\def\prdj#1{{\it Phys. Rev.} {\bf D{#1}}}
\def\npbj#1{{\it Nucl. Phys.} {\bf B{#1}}}

\def\plbj#1{{\it Phys. Lett.} {\bf B{#1}}}

\def\prepj#1{{\it Phys. Rep.} {\bf {#1}}}

\def\MPL #1 #2 #3 {Mod.~Phys.~Lett.~{\bf#1},\  #2 (#3)}
\def\NPB #1 #2 #3 {Nucl.~Phys.~{\bf#1},\  #2 (#3)}
\def\PLB #1 #2 #3 {Phys.~Lett.~{\bf#1},\  #2 (#3)}
\def\PR #1 #2 #3 {Phys.~Rep.~{\bf#1},\ #2 (#3)}
\def\PRD #1 #2 #3 {Phys.~Rev.~{\bf#1},\  #2 (#3)}
\def\PRL #1 #2 #3 {Phys.~Rev.~Lett.~{\bf#1},\  #2 (#3)}
\def\RMP #1 #2 #3 {Rev.~Mod.~Phys.~{\bf#1},\  #2 (#3)}
\def\ZP #1 #2 #3 {Z.~Phys.~{\bf#1},\  #2 (#3)}
\def\IJMP #1 #2 #3 {Int.~J.~Mod.~Phys.~{\bf#1},\  #2 (#3)}

\def\ibid{{\it ibid.}}
\def\caln{{\cal N}}
\def\cald{{\cal D}}
\def\DM{D$^-$}
\def\DP{D$^+$}
\def\NSM{NS$^-$}
\def\NSP{NS$^+$}
\def\HSM{HS$^-$}
\def\HSP{HS$^+$}

\def\leff{L_{\rm eff}}
\def\sign{{\rm sign}}
\def\br{B}
\def\breff{\br_{\rm eff}}
\def\wtil{\widetilde}

\def\zstar{Z^{\star}}

\def\rts{\sqrt s}
\def\eps{\epsilon}

\def\eg{{\it e.g.}}

\def\epem{e^+e^-}
\def\mupmum{\mu^+\mu^-}

\def\taup{\tau^+}
\def\taum{\tau^-}
\def\tauptaum{\taup\taum}

\def\lsim{\mathrel{\raise.3ex\hbox{$<$\kern-.75em\lower1ex\hbox{$\sim$}}}}
\def\gsim{\mathrel{\raise.3ex\hbox{$>$\kern-.75em\lower1ex\hbox{$\sim$}}}}
\def\@versim#1#2{\vcenter{\offinterlineskip
        \ialign{$\m@th#1\hfil##\hfil$\crcr#2\crcr\sim\crcr } }}
\def\zstar{Z^\star}

\def\slash#1{#1\hskip-6pt/\hskip2pt}

\def\etmiss{\slash E_T}

\def\ie{{\it i.e.}}

\def\gam{\gamma}

\def\anti{\overline}

\def\fbi{~{\rm fb}^{-1}}

\def\gev{\,{\rm GeV}}
\def\tev{\,{\rm TeV}}

\def\wt{\widetilde}

\def\rta{\rightarrow}
\def\mhalf{m_{1/2}}
\def\gl{\wt g}
\def\mgl{m_{\gl}}

\def\stopone{\wt t_1}

\def\sbottomone{\wt b_1}

\def\slep{\wt \ell}

\def\mslep{m_{\slep}}

\def\slepr{\wt \ell_R}
\def\mslepr{m_{\slepr}}

\def\hl{h^0}
\def\hh{H^0}
\def\ha{A^0}
\def\hp{H^+}
\def\hm{H^-}
\def\hpm{H^{\pm}}
\def\mhl{m_{\hl}}
\def\mhh{m_{\hh}}
\def\mha{m_{\ha}}
\def\mhp{m_{\hp}}
\def\mhpm{m_{\hpm}}
\def\tanb{\tan\beta}
\def\cotb{\cot\beta}
\def\mt{m_t}
\def\mb{m_b}
\def\mz{m_Z}
\def\mw{m_W}
\def\mgut{M_U}

\def\wp{W^+}
\def\wm{W^-}
\def\wpm{W^{\pm}}

\def\chitil{\wt\chi}
\def\cnone{\wt\chi^0_1}
\def\cntwo{\wt\chi^0_2}
\def\snu{\wt\nu}

\def\mcnone{m_{\cnone}}
\def\mcntwo{m_{\cntwo}}

\def\cpone{\wt \chi^+_1}
\def\cmone{\wt \chi^-_1}
\def\cpmone{\wt \chi^{\pm}_1}
\def\mcpone{m_{\cpone}}
\def\mcpmone{m_{\cpmone}}
\def\stauone{\wt \tau_1}
\def\mstauone{m_{\stauone}}

\documentstyle[12pt,equations]{article}
\def\ie{{\it i.e.}}

\def\9{\phantom 0}      
\renewcommand\linebreak{\unskip\break} 
\begin{document}
\input psfig.sty
\newlength{\captsize} \let\captsize=\small 
\newlength{\captwidth}                     

%
\font\fortssbx=cmssbx10 scaled \magstep2
\hbox to \hsize{
%
%
$\vcenter{
\hbox{\fortssbx University of California - Davis}
}$
\hfill
$\vcenter{
\hbox{\bf UCD-96-24} 
\hbox{\bf MADPH-96-969} 
\hbox{October, 1996}
}$
}

%
\medskip
\begin{center}
\bf
Detecting and Studying {\boldmath $\epem\to \hh\ha,\hp\hm$} in the MSSM:
Implications of Supersymmetric Decays and Discriminating GUT Scenarios
\\
\rm
\vskip1pc
{\bf John F. Gunion}\\
\medskip
{\em Davis Institute for High Energy Physics}\\
{\em Department of Physics, University of California, Davis, CA 95616}\\
\medskip
{\bf James Kelly}\\
\medskip
{\em Davis Institute for High Energy Physics}\\
{\em Department of Physics, University of California, Davis, CA 95616}\\
{\it and} \\
{\em Department of Physics, University of Wisconsin, Madison, WI 53706}\\
\end{center}

\begin{abstract}
We demonstrate that supersymmetric decays, as typified
by the predictions of several GUT-scale boundary condition
choices, do not prevent detection of $\zstar\to\hh\ha,\hp\hm$,
at a $1\tev-4\tev$ $\epem$ or $\mupmum$ collider operating
at anticipated luminosity.
For much of parameter space the relative branching
ratios for various SUSY and non-SUSY decays can be measured with sufficient 
accuracy that different GUT-scale boundary condition choices can be
distinguished from one another at a very high confidence level.
\end{abstract}

\section{Introduction}

\indent\indent 
The minimal supersymmetric model
(MSSM) is widely regarded as the most attractive
extension of the Standard Model (SM).
The approximate unification of coupling constants that occurs in the MSSM
at an energy scale of a few times $10^{16}\gev$ 
\cite{langacker} suggests the
appropriateness of treating the MSSM in the context
of a grand unified (GUT) model, in which the supersymmetry breaking 
parameters have simple universal values at the unification scale, $\mgut$.
The GUT framework is especially compelling in
that electroweak symmetry breaking (EWSB) is easily 
induced at a scale $\sim \mz$ as the soft mass-squared
of the Higgs field that couples to the top quark is driven 
to small (sometimes negative) values
by the associated large Yukawa coupling
during evolution to low energy scales.
Thus, it is important to consider the implications of GUT scenarios
for the detection of the Higgs bosons of the MSSM and to 
determine the extent to which (and strategies by which)
Higgs boson decay branching fractions
can be measured with accuracy sufficient to constrain GUT models.

The Higgs sector of the MSSM is reviewed in Refs.~\cite{hhg,dpfreport}.
The MSSM contains exactly two Higgs doublets, leading to two
CP-even Higgs bosons ($\hl$ and $\hh$, with $\mhl\leq \mhh$),
one CP-odd Higgs boson ($\ha$) and a charged Higgs pair ($\hpm$).
Crucial parameters for the Higgs sector are $\mha$ and $\tanb$
(the ratio of the vacuum expectation values for the neutral
Higgs fields that give mass to up-type and down-type quarks, respectively).
A fundamentally important GUT result is that essentially all
models with proper EWSB require $\mha> 200\gev$, with much larger
values being common.
This result has many important implications:
\begin{itemize}
\item The $\hl$ will be very SM-like, and, at fixed $\tanb$, will have a mass
near the upper bound predicted by including (two-loop/RGE-improved)
radiative corrections as computed for the known value of $\mt$ and 
the values for stop-squark masses and mixing predicted by the GUT.
For all scenarios considered (even those with $\mha$ well above a TeV),
$\mhl$ is below $\sim 130\gev$ and, as reviewed in Ref.~\cite{dpfreport},
will be discovered with relative ease at both the LHC and
any $\epem$ or $\mupmum$ collider with $\rts\gsim 500\gev$.
However, because the $\hl$ will be very SM-like, it will be quite
difficult to establish on the basis of precision measurements
that it is the MSSM $\hl$ and not the SM Higgs, especially if
$\mha\gsim 300-400\gev$ \cite{dpfreport}.
\item The $\hh$, $\ha$ and $\hpm$ will be approximately degenerate
in mass and will decouple from the vector boson sector. 
The coupling of the $\ha$ to $b\anti b$ [$t\anti t$] is given by
$\gamma_5$ times $-gm_b/(2\mw)\tanb$ [$-gm_t/(2\mw)\cotb$].
For large $\mha$, the couplings of
the $\hh$ asymptote to $i$ times these same coefficients.
The $\hp\to t\anti b$ coupling 
is proportional to $ig/(\sqrt 2\mw)(\mb P_R\tanb+\mt P_L\cotb)$.
\item
In most GUT scenarios, the high masses predicted for the $\hh$ and $\ha$
imply that decays to pairs of supersymmetric particles
will be important when $\tanb$ is not large and
$t\anti t$ decays are not kinematically allowed.  
For small to moderate $\tanb$ and $\mhh,\mha\gsim 2\mt$,
$t\anti t$ is the dominant mode unless the mass of
the lightest stop squark, $\stopone$, is small enough that 
decays to $\stopone\stopone$ are kinematically allowed. (This
does not happen in the GUT models we consider.) When $\tanb$ is large,
the enhanced $b\anti b$ coupling of the $\ha$ and $\hh$ imply
that $b\anti b$ decays will become dominant, even when SUSY and/or
$t\anti t$ decay modes are allowed. In the case of the $\hpm$,
SUSY decays always compete with the larger $tb$ decay mode since 
$\mhpm>\mt+\mb$ for the GUT scenarios considered. (In the GUT models
we consider, $\stopone \wtil b_1$ decays are not kinematically allowed.)
\item For $\mha\gsim 200\gev$ it is entirely possible that 
none of these heavy Higgs bosons could
be detected at the LHC (see the review of Ref.~\cite{dpfreport}),
even assuming the absence of SUSY decays. In terms of
the $(\mha,\tanb)$ parameter space plane, heavy Higgs discovery
is not possible once $\mha\gsim 200\gev$ if $\tanb\gsim 3$
and if $\tanb$ lies below an upper limit that increases with
increasing $\mha$ (reaching $\tanb\sim 15$ by $\mha\sim 500\gev$,
for example). More than likely,
the $\tanb\lsim 3$ discovery region would be diminished 
after including the SUSY decays of the $\hh$ and $\ha$
that are predicted to be important. Detection of the $\hh$ and
$\ha$ via such SUSY decays at the LHC appears to be very difficult
except in rather special situations.
\item The only large rate production modes for these heavy
Higgs bosons at an $\epem$ or $\mupmum$
collider will be $\zstar\to \hh\ha$ and $\zstar\to \hp\hm$.  These
modes are kinematically limited to $\mha\sim\mhh\sim\mhpm\lsim \rts/2$.
In particular, at a first $\epem$ collider
with $\rts=500\gev$ and $L=50\fbi$ observation is restricted to
roughly $\lsim 220-230\gev$, implying
that detection would not be possible in most GUT scenarios
\item Although single $\hh$ and $\ha$ production is significant
at a $\gam\gam$ collider facility for 
masses $\lsim 0.8\rts$, \ie\ about $400\gev$ at a $\rts=500\gev$
$\epem$ collider, backgrounds are such that very high luminosities
are required for discovery \cite{gkgamgamsusy} 
--- $L\gsim 200\fbi$ is required when either SUSY decays
are significant or $\tanb$ is large.
\end{itemize}

In combination, these results imply that $\hh$, $\ha$, and $\hpm$
detection may very well require employing the $\zstar\to \hh\ha$
and $\zstar\to\hp\hm$ production modes at 
an $\epem$ or $\mupmum$ collider with $\rts$ substantially above 500 GeV. 
Even if the $\hh$ and $\ha$ are observed at the LHC, studying
their decays and couplings would be much simpler in the pair modes.
Various aspects of Higgs pair production are discussed in Ref.~\cite{kalzer},
which appeared as we were completing the present work.

Our first goal is to determine the luminosity required
to guarantee observability of the $\zstar\to\hh\ha,\hp\hm$ 
modes regardless of the SUSY-GUT
decay scenario. We will consider collider energies of $1\tev$
and $4\tev$ (the latter being actively considered
\cite{mupmumconferences} for $\mupmum$ colliders), with integrated
luminosities up to $200\fbi$ and $1000\fbi$, respectively.
Our second goal will be to develop strategies for organizing
the rates observed for physically distinct final states so as to
yield information regarding
the relative branching fractions of different types of decay modes,
and to assess the extent to which such information can determine
the GUT scenario and its parameters given the expected experimental errors.

We find that if the integrated luminosity at $1\tev$ ($4\tev$) is close to
$200\fbi$ ($1000\fbi$) then detection
of the $\hh\ha$ and $\hp\hm$ pair production processes will be possible
over almost all of the kinematicaly allowed parameter space 
in the models we consider,
but that significant reductions in these luminosities will imply
gaps in parameter space coverage. A measurement
of the mass $\mha\sim\mhh\sim\mhpm$ already provides critical constraints
on the GUT model.  The correlation between this mass and the masses of
the charginos, neutralinos, and/or sleptons 
(as measured in direct production) determines 
the GUT scale boundary conditions (provided there is universality
for the standard soft-SUSY-breaking parameters)
and a fairly unique location in the parameter space of the GUT model
so singled out. In particular, $\tanb$ is determined.
Assuming full luminosity,
the relative Higgs branching fractions can be used to cross check the
consistency of the GUT model and confirm the parameter space location
with substantial precision.  For example,
the relative branching fractions for the $\hh,\ha,\hpm$ to decay to SUSY pair
particle states vs. Standard Model pair states provide a surprisingly
accurate determination of $\tanb$ given a measured value for $\mha$.
This $\tanb$ value must agree with that determined from the masses.
Other relative branching fractions provide complementary information
that can be used to further constrain the
GUT model, and can provide a determination of the
sign of the Higgs superfield mixing parameter.
Thus, a relatively thorough study of the full Higgs sector of the MSSM
will be possible and will provide consistency checks
and constraints that could single out the correct GUT model.

The organization of the paper is as follows.  In the next section,
we describe the six GUT models that we consider, and delineate
the allowed parameter space for each.  Contours of
constant Higgs boson, neutralino and chargino masses are given
within the allowed parameter space, and Higgs boson decay branching
fractions are illustrated. In Section 3, we demonstrate
that, for expected integrated luminosities at $\epem$ or $\mupmum$
colliders, detection of Higgs pair production will be possible
in final state modes where both Higgs bosons decay to final states
containing only $b$ or $t$ quarks, even though the branching fractions
for such final states are decreased due to competition from the
SUSY decay channels. Event rate contours as a function
of parameter space location are presented for the six GUT models.
In Section 4, we determine the prospects for measuring
the branching fractions for various Higgs boson decays, including
those for specific supersymmetric (SUSY) sparticle pairs.  The ability
to discriminate between different GUT models and to determine
the parameter space location within the correct GUT model on
the basis of Higgs decays is delineated. Section 5 summarizes
our results and conclusions.

\section{The GUT Models, Masses and Higgs Decays}

\indent\indent
In the simplest GUT treatments of the MSSM, soft supersymmetry breaking
at the GUT scale is specified by three universal parameters:
\begin{itemize}
\item $m_0$: the universal soft scalar mass;
\item $\mhalf$: the universal soft gaugino mass;
\item $A_0$: the universal soft Yukawa coefficient.
\end{itemize}
The absolute value of $\mu$ (the Higgs mixing parameter) is 
determined by requiring
that radiative EWSB gives the exact value of $\mz$
for the experimentally measured value of $\mt$; however,
the sign of $\mu$ remains undetermined. Thus,
the remaining parameters required to completely fix the model are
\begin{itemize}
\item $\tanb$: the vacuum expectation value ratio; and
\item sign($\mu$).
\end{itemize}
We remind the reader that a universal gaugino mass at the GUT scale
implies that $M_3:M_2:M_1\sim 3:1:1/2$ at scale $\sim\mz$. For models
of this class one also finds that $|\mu|\gg M_{1,2}$. These two facts
imply that the $\cnone$ is mainly bino, while $\cntwo$ and $\cpone$
are mainly wino, with heavier gauginos being mainly higgsino \cite{ghino}.
The running gluino mass $\mgl(\mgl)$
is roughly three times as large as $\mcntwo\sim\mcpone$
which in turn is of order twice as large as $\mcnone$. (The pole
gluino mass is generally substantially larger than $\mgl(\mgl)$
when squark masses are large.)

We will consider three representative GUT scenarios characterized by
increasingly large values of $m_0$ relative to $\mhalf$ (which
translates into increasingly large slepton masses as compared to
$\mcnone$, $\mcntwo$, and $\mcpone$):
\begin{itemize}
\item ``No-Scale'' (NS) \cite{noscale}: $A_0=m_0=0$;
\item ``Dilaton'' (D) \cite{dilaton}: $\mhalf=-A_0=\sqrt3 m_0$;
\item ``Heavy-Scalar'' (HS): $m_0=\mhalf$, $A_0=0$.
\end{itemize}
Within any one of these three scenarios, the model is completely specified
by values for $\mhalf$, $\tanb$ and sign($\mu$). We will present results
in the $(\mhalf,\tanb)$ parameter space for a given sign($\mu$) and
a given choice of scenario. Our notation will be $NS^-$ for the No-Scale
scenario with $\sign(\mu)<0$, and so forth.

\begin{figure}[[htbp]
\vskip 1in
\let\normalsize=\captsize   
\begin{center}
\centerline{\psfig{file=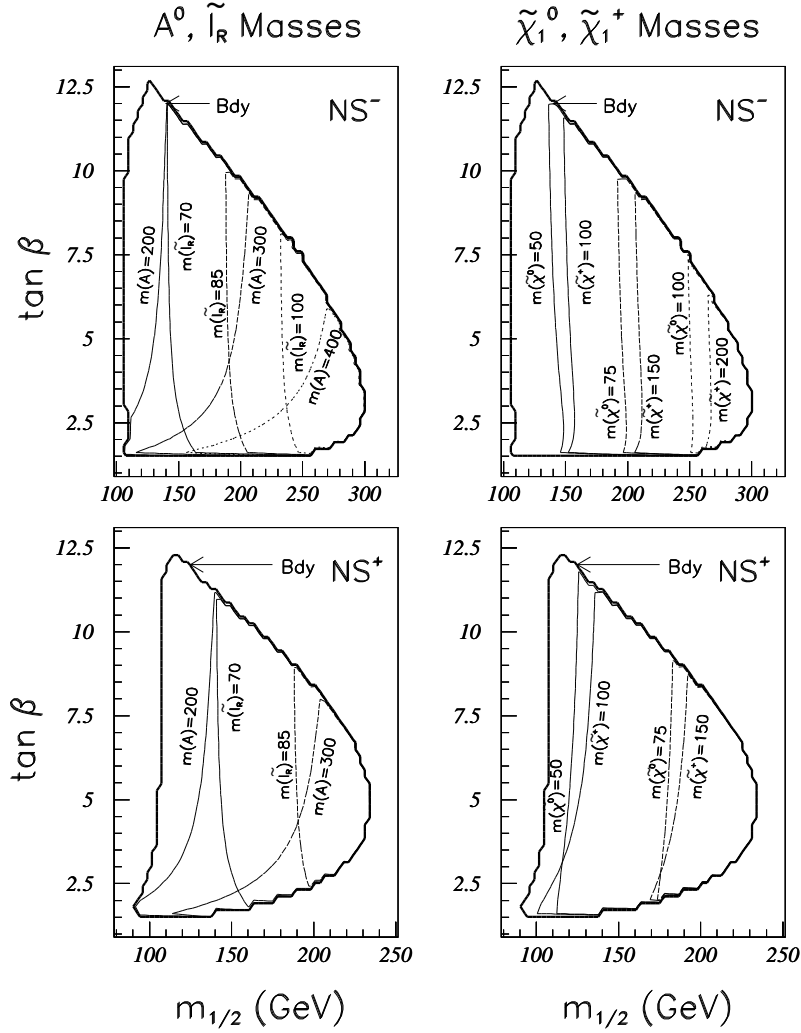,width=10.5cm}}
\smallskip
\begin{minipage}{12.5cm}       
\caption{
We show the $(\mhalf,\tanb)$ parameter space regions (bold outer
perimeter) within which 
we find a consistent EWSB solution for the No-Scale model.
Contours of constant mass are shown within the allowed region
for the $\cnone$, $\cpmone$, $\ha$ and $\slepr$. Results
for both signs of $\mu$ are shown.
}
\label{massesns}
\end{minipage}
\end{center}
\end{figure}

\begin{figure}[[htbp]
\vskip 1in
\let\normalsize=\captsize   
\begin{center}
\centerline{\psfig{file=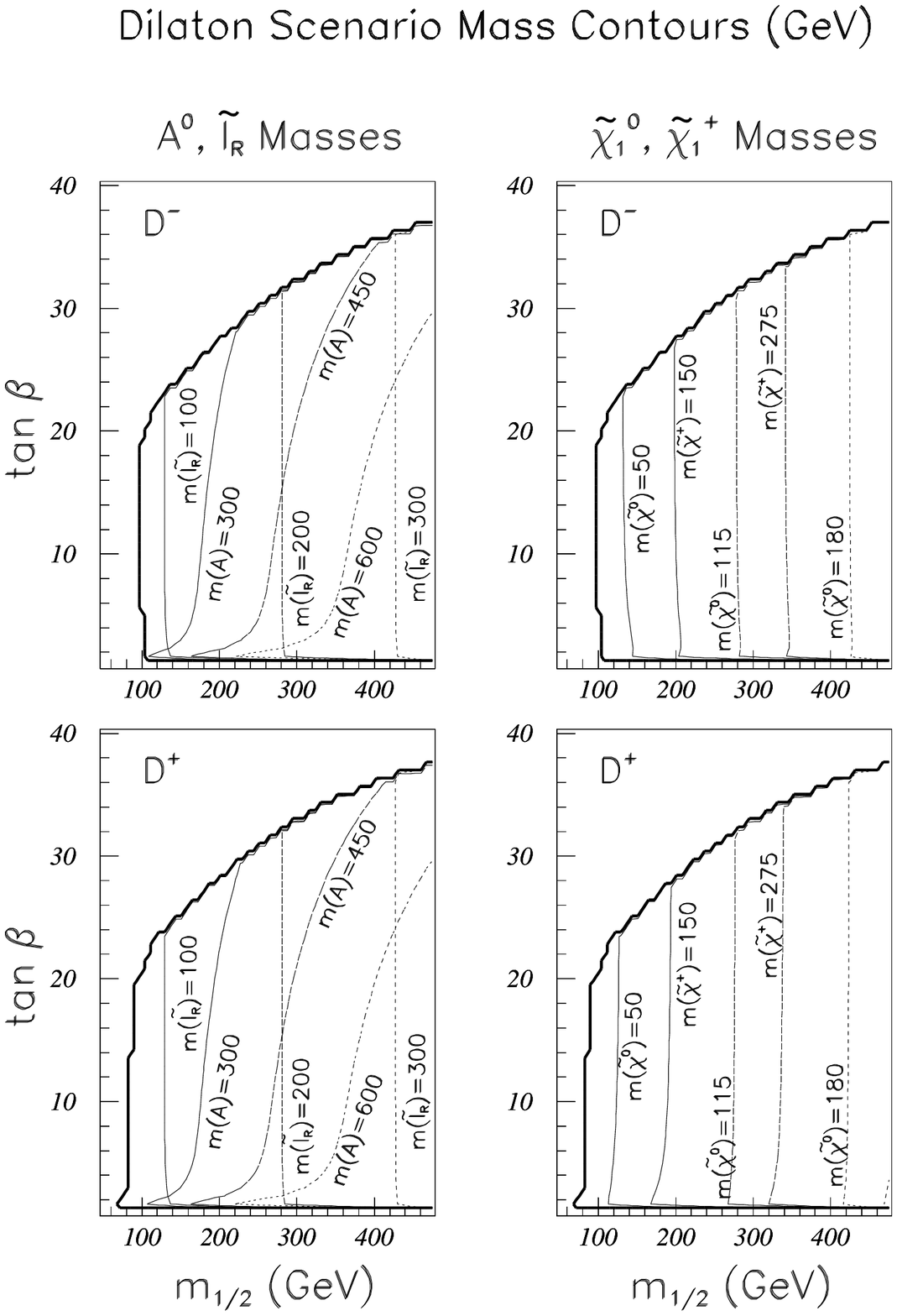,width=10.5cm}}
\smallskip
\begin{minipage}{12.5cm}       
\caption{
We show the $(\mhalf,\tanb)$ parameter space regions (bold outer
perimeter) within which 
we find a consistent EWSB solution for the Dilaton model.
Contours of constant mass are shown within the allowed region
for the $\cnone$, $\cpmone$, $\ha$ and $\slepr$. Results
for both signs of $\mu$ are shown.
}
\label{massesd}
\end{minipage}
\end{center}
\end{figure}

\begin{figure}[[htbp]
\vskip 1in
\let\normalsize=\captsize   
\begin{center}
\centerline{\psfig{file=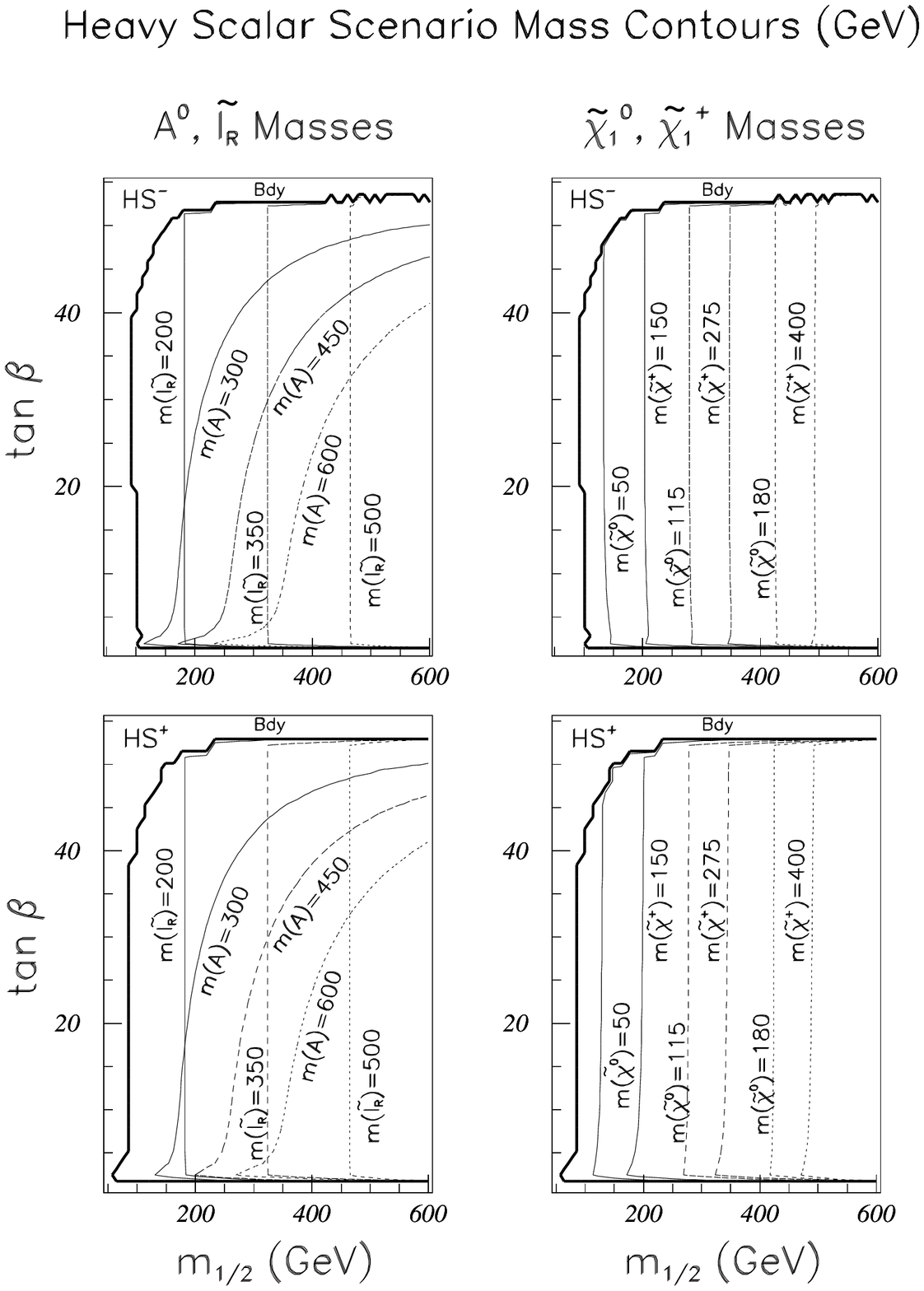,width=10.5cm}}
\smallskip
\begin{minipage}{12.5cm}       
\caption{
We show the $(\mhalf,\tanb)$ parameter space regions (bold outer
perimeter) within which 
we find a consistent EWSB solution for the Heavy-Scalar model.
Contours of constant mass are shown within the allowed region
for the $\cnone$, $\cpmone$, $\ha$ and $\slepr$. Results
for both signs of $\mu$ are shown.
}
\label{masseshs}
\end{minipage}
\end{center}
\end{figure}

In Figures~\ref{massesns}, \ref{massesd}, and 
\ref{masseshs} we display the 
allowed $(\mhalf,\tanb)$ parameter space for the NS, D and HS scenarios,
respectively.
The boundaries of the allowed parameter space are fixed by
experimental and theoretical constraints as follows:
\begin{itemize}
\item The left-hand boundary at low $\mhalf$ derives from
requiring that $Z\to$SUSY decays not violate LEP1 limits.
\item The low-$\tanb$ boundary is obtained by requiring that
the $t$-quark Yukawa coupling remain perturbative in evolving
from scale $\mz$ to the GUT scale.
\item In the NS scenario, the allowed parameter space is finite
by virtue of two competing requirements.
First, there is an upper bound on $\tanb$ as a function
of $\mhalf$ obtained by requiring that the LSP (always 
the $\cnone$ in the allowed region) not be charged
(\ie\ we require $\mstauone\geq\mcnone$).\footnote{This bound 
is especially strong in the NS
scenario due to the fact that $m_0=0$ 
implies very modest masses for the sleptons, in
particular the $\stauone$, at scale $\mz$.}
Second, there is the lower bound on $\tanb$ required
by $t$-quark Yukawa perturbativity. One finds that
for large enough $\mhalf$ the upper bound drops below the lower bound.
\item The upper bound on $\tanb$ as a function
of $\mhalf$ in the D scenario comes from demanding that the LSP
not be charged.
\item In the HS scenario, the upper bound on $\tanb$ arises
by requiring that the SM-like light Higgs mass lie above the current
limit of $\mhl\gsim63\gev$. (In the HS scenario, for fixed $\mhalf$,
$\mha$ becomes smaller and smaller as  $\tanb$ increases until eventually
it approaches zero forcing $\mhl$ to decline rapidly. In other scenarios,
with lighter scalar masses and hence sleptons, 
the LSP becomes charged before $\tanb$ becomes
so large that $\mha$ starts declining rapidly.)
\item In the D and HS scenarios, there is no upper bound on $\mhalf$
unless cosmological constraints are imposed. High $\mhalf$
values (roughly, $\mhalf\gsim 500\gev$ \cite{ganderson}) are, however,
disfavored by naturalness considerations.
\end{itemize}

Before proceeding, we provide a few technical notes.
First, we note that the evolution equations
must be implemented very carefully when considering very large $\mha$
values.  In order to avoid instabilities\footnote{Such instabilities
are found, for example, in ISASUGRA \cite{isasugra}.}
deriving from unnaturally large (and hence unreliable) one-loop corrections
(for going from running masses to pole masses),
we found it necessary to terminate evolution for soft
masses at scales of order the associated final physical squark, slepton
and heavy Higgs masses. In this way, the one-loop corrections are kept small
and the physical masses obtained are reliable. The evolution program
we employed is based on one developed by C.~H. Chen \cite{chen}.
Results at low mass scales were checked against results obtained
using the programs developed for the work of Refs.~\cite{bgkp}
and \cite{gkgamgamsusy}.
Once the appropriate low-energy parameters were determined from
the evolution, we then employed ISASUSY \cite{isasugra} 
to obtain the branching
ratios for the Higgs boson and subsequent chain decays. The ISASUSY results
were cross-checked with our own programs. The decay results
were then combined with Higgs boson pair production rates to determine
rates for specific classes of final states.

\subsection{Sparticle and Higgs Masses}

\indent\indent
Also displayed in Figs.~\ref{massesns}, \ref{massesd}, and \ref{masseshs}
are contours of constant $\mcnone$, $\mcpone$, $\mslepr$, and $\mha$.
These reveal the importance of detecting 
the heavy Higgs bosons and measuring their masses accurately.
The masses of the inos and the sleptons will presumably be
measured quite accurately, and the figures show
that they will determine in large measure
the values of $\mhalf$ and $m_0$. But the rather vertical nature
of the $\mcnone$, $\mcpmone$, and $\mslepr$ contours implies
that $\tanb$ is likely to be poorly determined
from these masses alone. Fortunately, the $\mha$ contours are not
nearly so vertical, implying that a measurement of $\mha$
can be combined with the $\mhalf$ determination from the ino masses
to fix a value of $\tanb$.  The accuracy of this determination depends
upon the accuracy with which $\mha$ (and $\mhh$, $\mhpm$) can be measured.
For discovery in the $\ha\to b\anti b$ decay mode (as possible
for almost all model parameter choices at full luminosity, see later),
this accuracy is fixed by the $b\anti b$ mass resolution.  At an $\epem$
collider, a resolution of $\pm\Delta M_{bb}\sim \pm10\gev$ 
is probably attainable.  For a 
large number, $N$, of events, $\mha$ can be fixed to a value of order
$\Delta M_{bb}/\sqrt N$, which for $N=20$ (our minimal
discovery criterion) would imply $\Delta \mha\sim 2-3\gev$. Examination
of the figures shows that such mass uncertainty will lead to a rather
precise $\tanb$ determination within a given GUT model, except
at low $\mha$ and high $\tanb$ in the NS case.

\subsection{Higgs Decays}

\indent\indent
Let us now turn to the decays of the heavy Higgs bosons of the MSSM.
As already noted, our ultimate goal is to use these
to confirm/re-enforce the correctness of both
the model and the parameter choices within the model that has
been singled out by the mass measurements.
The most important common feature of the GUT models we consider is that
squarks are always sufficiently heavy that
decays of Higgs bosons to squark pairs are not kinematically allowed.
This is true even for the NS boundary conditions with $m_0=0$,
in which the large squark masses derive from the substantial evolution
of the colored soft-scalar masses to positive values
as the scale decreases from $\mgut$ towards $\mz$.
In order that the squarks be light enough for squark pairs to appear
in Higgs decays, substantial breaking of the universality of 
soft-SUSY-breaking scalar masses at the GUT scale is required.
For example, light sbottom and stop squarks can be consistent
with radiative EWSB via evolution 
if the Higgs soft scalar masses are much larger
than the squark (in particular, stop and sbottom) 
soft scalar masses at $\mgut$.  
In this case, $\hh,\ha\to \stopone\stopone, \sbottomone\sbottomone$ 
and $\hp\to \stopone\sbottomone$ pair channels would dilute
the SM decay modes of the Higgs to a much greater extent than do
the ino and slepton decays 
in the models discussed here. Strategies for detecting
and studying $\hh\ha$ and $\hp\hm$ pair production would have to
be reconsidered. In any case, there would be no difficulty
in distinguishing models with light stop and/or sbottom squarks
from the NS, D and HS models considered here.

\begin{figure}[[htbp]
\vskip 1in
\let\normalsize=\captsize   
\begin{center}
\centerline{\psfig{file=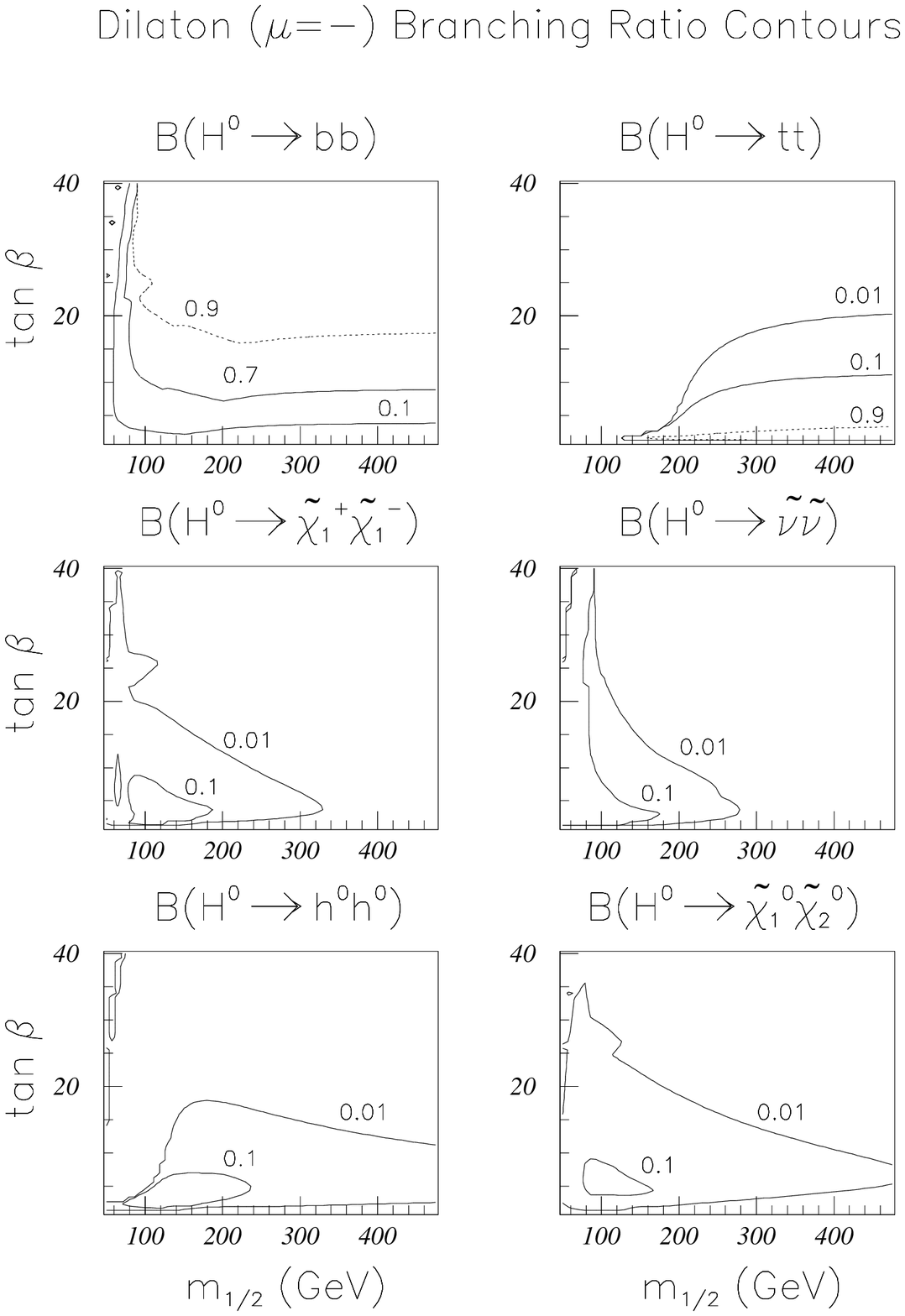,width=10.5cm}}
\smallskip
\begin{minipage}{12.5cm}       
\caption{a)
We show contours within the $(\mhalf,\tanb)$ parameter space
of constant branching fraction for the
$\hh\to b\anti b$, $t\anti t$, $\cpone\cmone$, $\snu\snu$,
$\hl\hl$, and $\cnone\cntwo$ decay channels.
Results are for the \DM\ scenario.
}
\label{decaysd}
\end{minipage}
\end{center}
\end{figure}
\addtocounter{figure}{-1}

\begin{figure}[[htbp]
\vskip 1in
\let\normalsize=\captsize   
\begin{center}
\centerline{\psfig{file=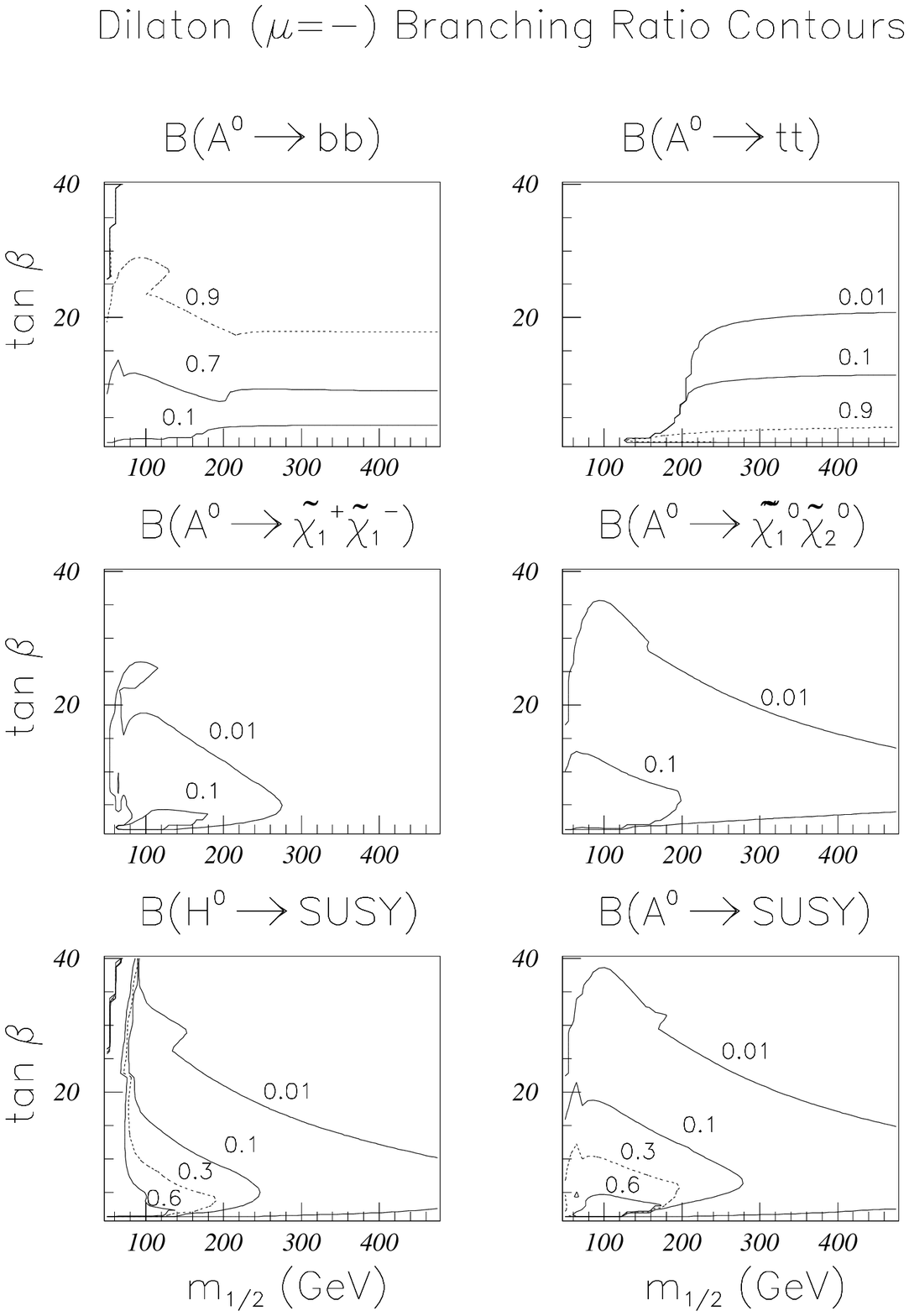,width=10.5cm}}
\smallskip
\begin{minipage}{12.5cm}       
\caption{b)
We show contours within the $(\mhalf,\tanb)$ parameter space
of constant branching fraction for the
$\ha\to b\anti b$, $t\anti t$, $\cpone\cmone$,
and $\cnone\cntwo$ decay channels, as well as for $\hh\to {\rm SUSY}$
and $\ha\to {\rm SUSY}$, summed over all SUSY channels.
Results are for the \DM\ scenario.
}
\end{minipage}
\end{center}
\end{figure}
\addtocounter{figure}{-1}

\begin{figure}[[htbp]
\vskip 1in
\let\normalsize=\captsize   
\begin{center}
\centerline{\psfig{file=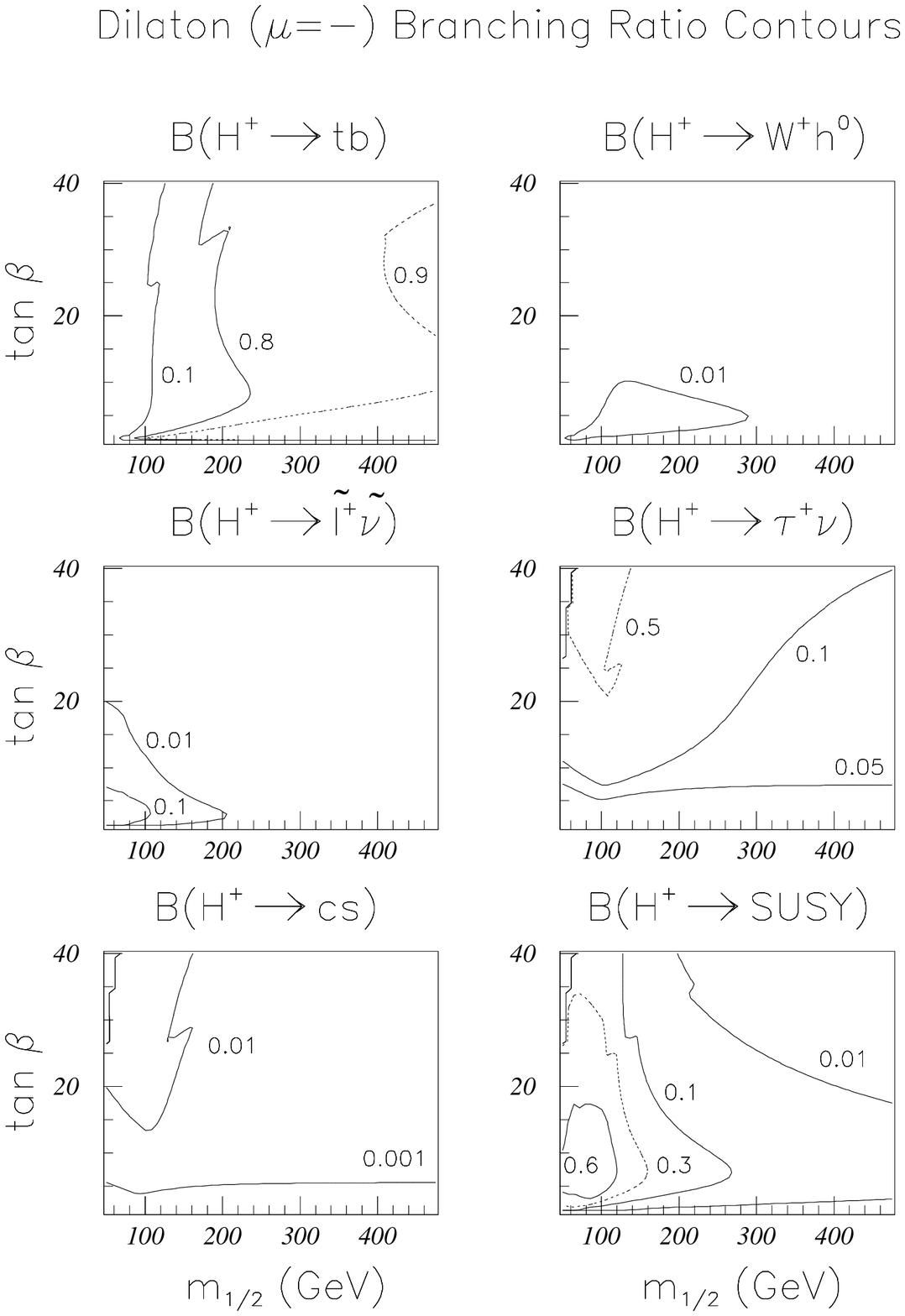,width=10.5cm}}
\smallskip
\begin{minipage}{12.5cm}       
\caption{c)
We show contours within the $(\mhalf,\tanb)$ parameter space
of constant branching fraction for the
$\hp\to t\anti b$, $\wp\hl$, ${\slep}^+\snu$, $\taup\nu$, $c\anti s$,
and SUSY, summed over all SUSY channels.
Results are for the \DM\ scenario.
}
\end{minipage}
\end{center}
\end{figure}

In fact, the three models we consider are rather similar
to one another in most respects. Thus, they provide
a good testing ground for assessing the extent to which
we can distinguish between models 
by using experimental information from the Higgs sector.
We shall see that Higgs branching
ratios depend substantially on the particular model choice and on
the precise location in parameter space within a given model.
Figures~\ref{decaysd}a, \ref{decaysd}b, and \ref{decaysd}c
illustrate the dependence of Higgs branching fractions upon parameter
space location for the $\mu<0$ Dilaton (\DM) scenario.
In these figures, we give contours
of constant branching fractions for $\hh$, $\ha$ and $\hp$ decays.
The decay channels $b\anti b$, $t\anti t$, 
$\cpone\cmone$, $\cnone\cntwo$, and the sum over all SUSY decay
channels, are considered for both the $\hh$ and $\ha$. In addition,
we show the $\hl\hl$ and $\snu\snu$ (summed over all $\snu$ types) 
branching fractions
for the $\hh$. (The $\snu\snu$ branching fraction for the $\ha$ is
very tiny.)  The $\ha\to Z\hl$ branching fraction is small,
but, as we shall see, measurable in some regions of parameter 
space. For the $\hp$ we display branching fraction contours for
$t\anti b$, $\wp\hl$, $\slep_L\snu$,
$\taup\nu$, $c\anti s$, and the sum over all SUSY decays.
[$\br(\ha\to Z\hl)$ is similar to $\br(\hp\to \wp\hl)$.]
Several important features of these plots deserve emphasis.
\begin{itemize}
\item For the $\hh$ and $\ha$, 
the net branching fraction for SUSY decays declines rapidly with
increasing $\tanb$ due to the enhancement of the $b\anti b$ coupling
and, hence, increasing relative importance of $b\anti b$ decays.
\item SUSY decays of the $\hh$ and $\ha$ are also small when
$\mhh,\mha>2\mt$, with the relative branching fraction 
$\br({\rm SUSY})/\br(t\anti t)$
saturating to a constant value below 0.1 for large $\mha$ (equivalently
large $\mhalf$) at fixed $\tanb$.
\item For $\mhh,\mha>2\mt$, the ratio of $b\anti b$ to $t\anti t$
branching fractions rises very rapidly as $\tanb$ increases.
\item The SUSY decay branching fraction of the $\hp$ is relatively independent
of $\tanb$ for lower $\mhalf$ values.
\item $\br(\hp\to \wp\hl)$ [as well as $\br(\ha\to Z\hl)$] 
is only signficant when $\tanb$ and $\mhalf$ are both small.
\item $\br(\hh\to\hl\hl)$ is significant for a larger range of 
modest $\tanb$ and $\mhalf$ values than the former two branching fractions.
\item $\br(\hp\to \taup \nu)$ remains significant ($\gsim 0.1$) for
a range of $\tanb$ values that becomes increasingly large as $\mhalf$
increases.
\end{itemize}
These figures show that a measurement of
several ratios of branching fractions (\eg\ SUSY/$b\anti b$
for the $\hh,\ha$ and SUSY/$t\anti b$ for the $\hp$)
would determine the values of $\tanb$ and $\mhalf$.
Branching ratios in the other five scenarios 
display a more or less similar pattern to that found in the \DM\
case, although the numerical values at any given $(\mhalf,\tanb)$
location can differ substantially.
For any given GUT scenario, definite predictions
for all other experimental observables are then possible and
could be checked for consistency with observations. In particular,
the predicted Higgs, neutralino, and chargino masses should agree with
the measured values if the GUT scenario is the correct one.

\section{Discovering the {\boldmath $\hh$, $\ha$ and $\hpm$}}

\indent\indent
In this section, we determine the luminosity required in order that
discovery of $\hh\ha$ and $\hp\hm$ be guaranteed over essentially
all of the allowed parameter space of the three scenarios.
For the models considered in this paper,
we find that discovery is always easiest by employing final
states in which neither of the Higgs bosons of the pair
decays to a final state containing SUSY particles.
The final state configurations we employ for discovery 
are listed below, along with techniques for isolating them
from backgrounds.
\begin{itemize}
\item I) $\hh\ha\to 4b$: We demand
observation of four jets which separate into two nearly equal mass
two-jet pairs. Event rates for this mode [labelled by $N(4b)$] include a factor
of $\br(\hh\to b\anti b)\br(\ha\to b\anti b)$.
\item II) $\hh\ha$ with $\hh\to \hl\hl\to 4b$ and $\ha\to X$: 
it would be sufficient to observe the two $\hl$'s by demanding
two jet pairs that reconstruct to the known $\mhl$ recoiling
against a reconstructed (from incoming energy and net $\hl\hl$ pair 
four momentum) `missing' mass that is the same as the $\hl\hl$ pair
mass. Event rates for this mode [labelled by $N(hh)$] include a factor
of $\br(\hh\to \hl\hl)[\br(\hl\to b\anti b)]^2$.
\item III) $\hh\ha\to 4t$: We can simply
demand $\geq 10$ visible (and moderately energetic/separated) leptons/jets.
The predicted rate for such states on the basis of QCD (including
$4t$ production) is quite small.  Because of inefficiencies
associated with combinatorics, we would not require
direct reconstruction of the $W$'s or  $t$'s (implying that
we would also not be able to require roughly
equal Higgs boson masses).
Event rates for this mode [labelled by $N(4t)$] include the effective branching
ratio for $\hh\ha\to \geq 10$ visible leptons/jets, given by 
$\br(\hh\to t\anti t)\br(\ha\to t\anti t)\br(t\anti t t\anti t\to\geq
10~{\rm visible})$.
\item IV) $\hp\hm\to 2t2b$: We insist on 8 jets or 
1 lepton plus 6 jets (in particular, fewer than 10 visible leptons/jets
so as to discriminate from the above $4t$ final states)
and possibly require that one $W$ and the associated $t$ be reconstructed.
Event rates for this mode [labelled by $N(tb)$] include a factor
of $[\br(\hp\to t\anti b)]^2\left\{2\br(t\to 2j b)\br(t\to \ell^+\nu b)
+[\br(t\to 2jb)]^2\right\}$.
\end{itemize}
There will also be an overall efficiency
factor for detector coverage and for
experimentally isolating and detecting these modes.  This
will be incorporated in our yearly event rate estimates
by reducing the total luminosity available (presumed
to be $L=200\fbi$ per year at $\rts=1\tev$ and $L=1000\fbi$ per year
at $\rts=4\tev$) by an overall efficiency factor of 40\% (to $\leff=80\fbi$
and $\leff=400\fbi$, respectively).
We have not performed a detailed simulation, but believe that
that such an efficiency is not unreasonable given the fact that 
backgrounds are relatively small for the above outlined signatures.
In particular, since all the final states contain at least four $b$ jets,
we can require one or two $b$-tags (in order to eliminate any residual
QCD background) without incurring significant penalty, given that
the vertex-tagger should have efficiency of 60\% or better
for any single $b$-jet within its acceptance.
(Tagging of $b$-jets would be desirable for
cleanly separating $\hh\ha$ from $\hp\hm$ final states.
In the absence of any $b$-tagging there would
be a small number of  $\hp\hm\to 4j$ (with $j=c,s$) events
that would combine with the $\hh\ha\to 4b$
final states to the extent that $\mhp\sim\mhh\sim\mha$.)
After including branching fractions and the 40\% efficiency,
something like 20 events should be adequate for detection. 
In our graphs, we will display 20, 50 and 200 event contours.

If the $\hh\ha\to 4t$ mode is dominant, we will wish to reconstruct
the mass of either the $\ha$ or the $\hh$ 
from the $4j2b$ decay mode of one of the $t\anti t$ pairs.
This will be important both
as a means for measuring the mass and also as a means for triggering
on $\hh\ha$ pair production using just one of the two members of the
pair (see Section 4). There will be a further
efficiency factor (on top of the above overall efficiency factor)
for isolating the relevant events and then reconstructing
the mass of the $\ha$ or $\hh$. We estimate this additional efficiency
factor be of order 25\% {\it each} for the $\ha$ and $\hh$. This is the result
that would be obtained from $[\br(t\to 2jb)]^2\hat\eps$, with
$\hat\eps=0.55$ for combinatoric and other problems.
The low net efficiency ($\sim 0.2=2\times \sim 0.25 \times 0.4$)
for events in which either $\mhh$ or $\mha$ could
be fully reconstructed implies
that an accurate determination of $\mhh\sim\mha$
would require several years of running if $\hh\ha\to 4t$ is the dominant
final state.

There are several reasons why non-SUSY final states are
best for discovery:
\begin{itemize}
\item As illustrated in Figs.~\ref{decaysd}a-c,
branching fractions for SM decays, \eg\ $\ha,\hh \to b\anti b$ or $t\anti t$
and $\hp\to t\anti b$, do not fall much below $0.1$;
\item Unlike the $b\anti b$ channel,
mass reconstruction in SUSY modes is not possible (due to missing
energy).
\item Particle multiplicities
in the $4t$ and $2t2b$ final states are sufficiently
large to be very distinctive and free of background, unlike many
of the  final states associated with SUSY decays.
\end{itemize}

\begin{figure}[[htbp]
\vskip 1in
\let\normalsize=\captsize   
\begin{center}
\centerline{\psfig{file=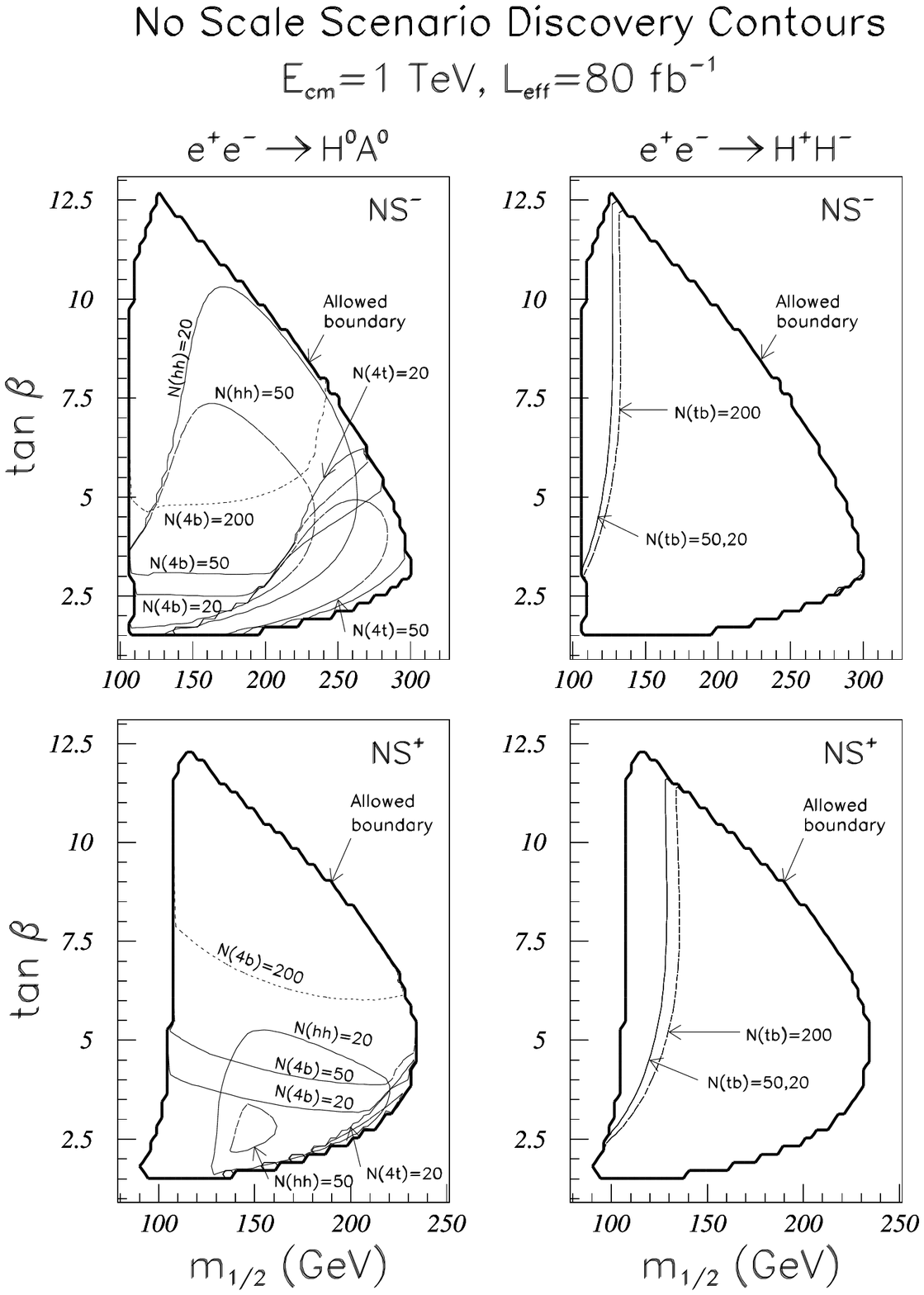,width=10.5cm}}
\smallskip
\begin{minipage}{12.5cm}       
\caption{
We show NS model 20, 50 and 200 event contours 
within the kinematically accessible
portion of the allowed $(\mhalf,\tanb)$ parameter space 
for $\hh\ha$ discovery modes I, II, III
and $\hp\hm$ discovery mode IV, assuming $\leff=80\fbi$.
}
\label{discns}
\end{minipage}
\end{center}
\end{figure}

\begin{figure}[[htbp]
\vskip 1in
\let\normalsize=\captsize   
\begin{center}
\centerline{\psfig{file=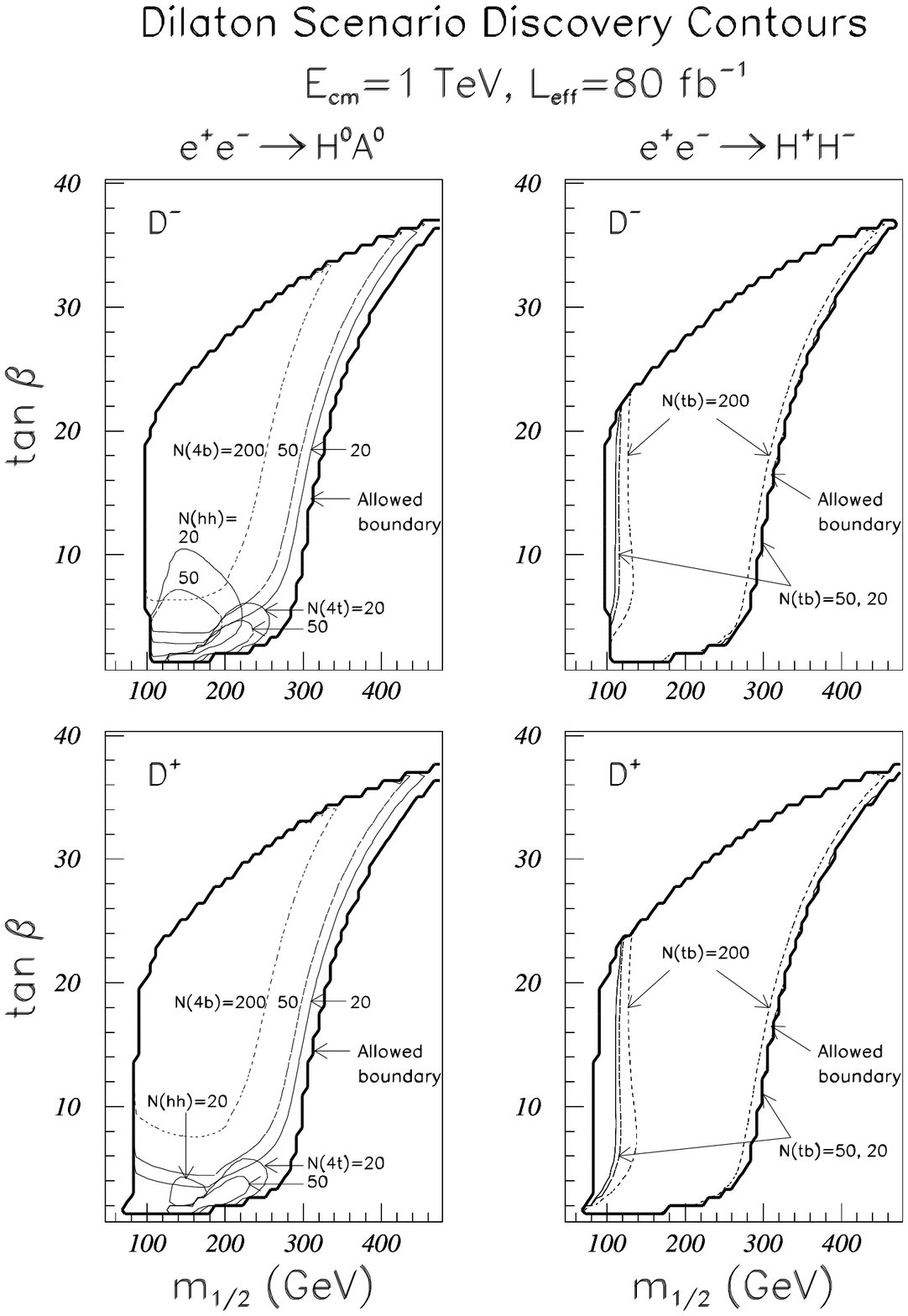,width=10.5cm}}
\smallskip
\begin{minipage}{12.5cm}       
\caption{
We show D model 20, 50 and 200 event contours 
within the kinematically accessible
portion of the allowed $(\mhalf,\tanb)$ parameter space 
for $\hh\ha$ discovery modes I, II, III
and $\hp\hm$ discovery mode IV, assuming $\leff=80\fbi$.
}
\label{discd}
\end{minipage}
\end{center}
\end{figure}

\begin{figure}[[htbp]
\vskip 1in
\let\normalsize=\captsize   
\begin{center}
\centerline{\psfig{file=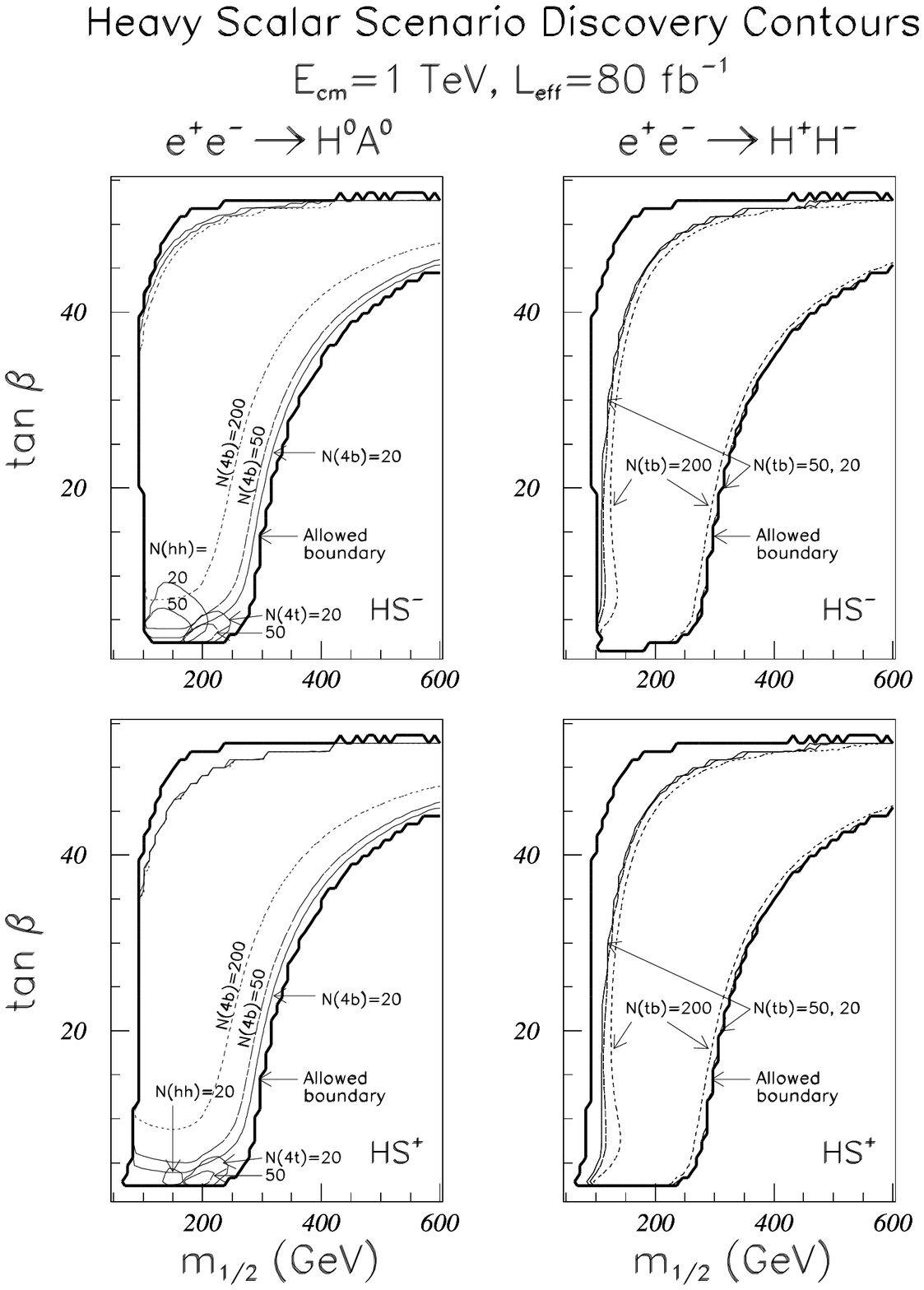,width=10.5cm}}
\smallskip
\begin{minipage}{12.5cm}       
\caption{
We show HS model 20, 50 and 200 event contours 
within the kinematically accessible
portion of the allowed $(\mhalf,\tanb)$ parameter space 
for $\hh\ha$ discovery modes I, II, III
and $\hp\hm$ discovery mode IV, assuming $\leff=80\fbi$.
}
\label{dischs}
\end{minipage}
\end{center}
\end{figure}

In Figures~\ref{discns}, \ref{discd}, and \ref{dischs}
(for the NS, D and HS scenarios, respectively)
we give the 20, 50 and 200 event contours in the
$(\mhalf,\tanb)$ parameter plane for $\hh\ha$
discovery modes I, II and III and $\hp\hm$ discovery
mode IV at $\rts=1\tev$. We assume $L=200\fbi$
and $\eps=40\%$ efficiency, \ie\ $L_{\rm eff}\equiv L\eps=80\fbi$.
Results are displayed for both signs of $\mu$.  Also shown 
are the boundaries defined by the kinematically accessible 
$\mhh+\mha\leq \rts$ or $2\mhpm\leq\rts$
portion of the allowed parameter space (bold solid lines).
In comparing scenarios, it will be important to note that the NS scenario
plots have greatly expanded axis scales relative to plots
for the D and HS scenarios.

As noted earlier, 20 events is likely to be adequate
for discovery; the 50 event contour at $\leff=80\fbi$ would
probably allow discovery at $\leff=32\fbi$ and the 200 event
contour would allow discovery at $\leff=8\fbi$.
These figures show that for all three GUT scenarios at least 20 
$\hh\ha$ events are present in one or more of the modes I-III throughout
almost the entire kinematically accessible
portion of the allowed $(\mhalf,\tanb)$ parameter space.
If the 50 event contours are appropriate
(because $\leff$ is a factor of 2.5 smaller) then one begins to see some,
but not enormous, sections of parameter space such that $\hh\ha$
detection would not be possible. If efficiencies and integrated
luminosity were in combination a factor of 10 worse than anticipated, 
the 200 event contours might apply; they 
indicate that $\hh\ha$ detection would then be
possible only in the part parameter space characterized by
small values of $\mhalf$ and large values of $\tanb$.

\begin{figure}[[htbp]
\vskip 1in
\let\normalsize=\captsize   
\begin{center}
\centerline{\psfig{file=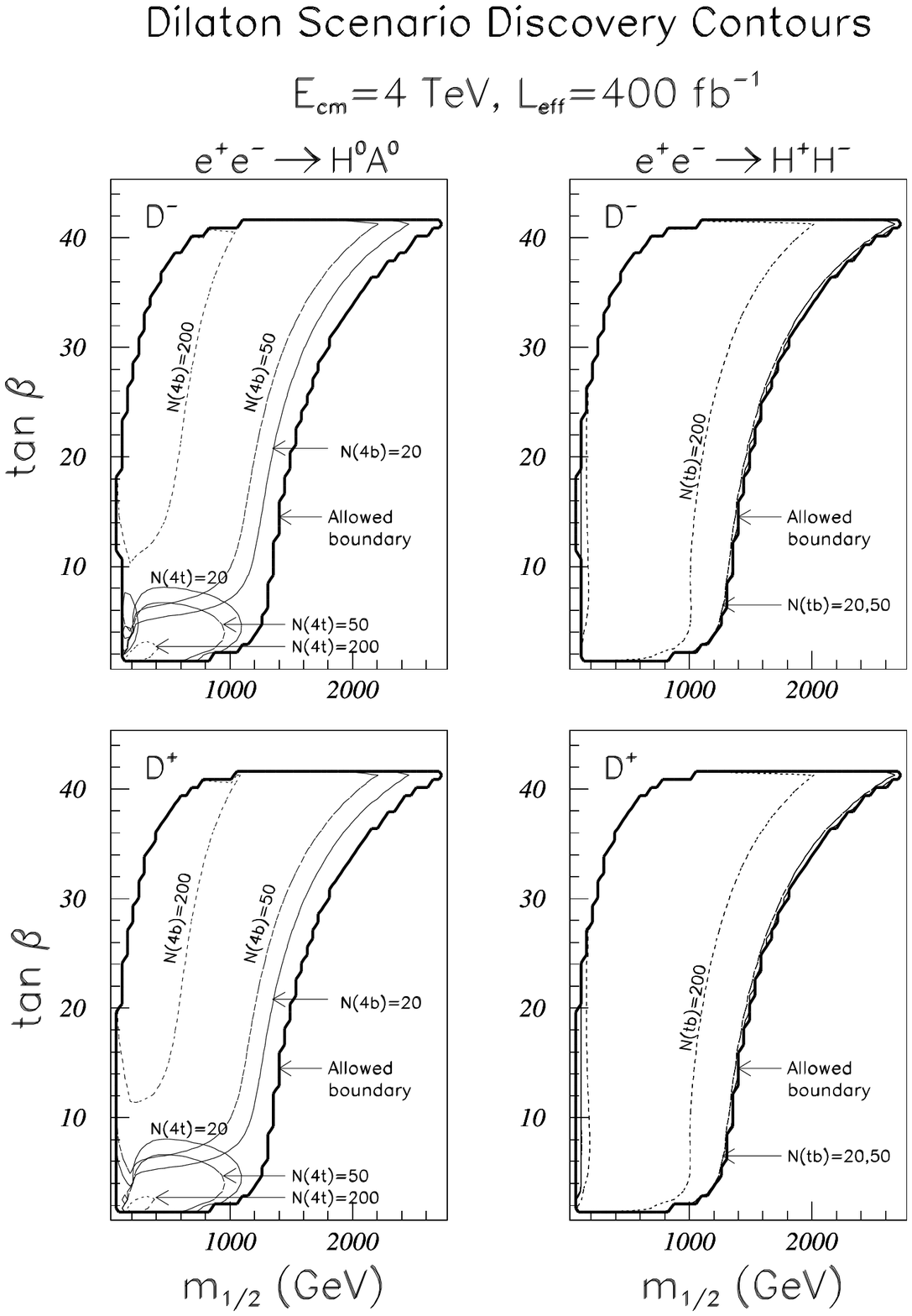,width=10.5cm}}
\smallskip
\begin{minipage}{12.5cm}       
\caption{
We show D model 20, 50 and 200 event contours 
within the kinematically accessible
portion of the allowed $(\mhalf,\tanb)$ parameter space 
for $\hh\ha$ discovery modes I, II, III
and $\hp\hm$ discovery mode IV, assuming $\leff =400\fbi$ at 
$\protect\rts =4\tev$.
}
\label{discdhi}
\end{minipage}
\end{center}
\end{figure}

In these same figures, the
20, 50 and 200 event contours for the $\hp\hm$ discovery
mode IV show that 20 events are found for all 
of the constraint and kinematically allowed parameter space
except a small wedge at small $\mhalf$ values. The 200 event contours 
(equivalent to 20 events at $\leff=5\fbi$) cover nearly as
a large section of parameter space.
Thus, even if efficiency and luminosity
are in combination a factor of 10 worse than anticipated, $\hp\hm$ discovery 
after just one year of running would be possible over the bulk 
of parameter space.
The somewhat better guarantees for the $\hp\hm$
mode as compared to the $\hh\ha$ mode derive simply from the larger
$\hp\hm$ cross section which is roughly a factor of 3 larger
than that for $\hh\ha$.

This same analysis can be repeated for $4\tev$ and $L=1000\fbi$
(implying $L=400\fbi$ for $\eps=0.4$ efficiency)
with very similar results.  The kinematic range
is much greater, allowing $\hh\ha$ and $\hp\hm$ production out
to masses $\mha\sim\mhh\sim\mhpm\lsim 2\tev$.
(The limited NS parameter space implies that
such energies are not needed were this the correct GUT scenario). 
If 20 events are
adequate (and they would certainly be rather spectacular events
at high Higgs boson mass) then both $\hh\ha$ and $\hp\hm$ detection
would be possible for nearly all of the constraint/kinematically
allowed parameter space for all three GUT scenarios.  
Dimunition in coverage due to poorer efficiency or
lower luminosity follows much the same pattern
as described for the $\rts=1\tev$, $L=200\fbi$ case. To illustrate,
we present the 20, 50 and 200 event contours for $\hh\ha$ (final
states I-III) and $\hp\hm$ (final state IV) in the D scenario,
Fig.~\ref{discdhi}.

\section{Measuring Ratios of Branching Fractions and Discriminating
Between Models}

\indent\indent
In this section we discuss the prospects for measuring
the relative size of the various branching fractions for different
decay modes of a given Higgs boson and for using
such measurements to pin down the GUT model
and parameter choices within a given GUT model.  Additional
information is contained in the absolute rates for different types
of final states. However, it is likely that greater uncertainty
will be associated with absolute rates than with ratios of rates,
since some types of efficiencies will cancel out of the ratios.

The key to determining the relative magnitude of the branching fractions
for different final state decays is to first identify 
and mass-reconstruct (`tag') one of the Higgs bosons in the $\hh\ha$
or $\hp\hm$ pair final state, and then compare the relative
rates for different types of decays of the second Higgs boson.
Identification and mass-reconstruction of the first Higgs boson
requires using one of its fully reconstructable final states.
As additional verification that the event corresponds
to Higgs pair production, we would require that the missing mass
(as computed using the incoming center-of-mass four-momentum
and the four-momentum of the reconstructed Higgs)
be roughly equal to the mass of the identified Higgs. 
For identification and mass-reconstruction of the first Higgs boson,
we employ:
\begin{itemize}
\item $\hh\ha$ with $\hh\to 2b$ or $\ha\to 2b$;
\item $\hh\ha$ with $\hh\to 2t$ or $\ha\to 2t$ ---
note that, unlike the $4t$ discovery channel, reconstruction
of the $2t$ mass will be necessary, and will
be accompanied by an extra efficiency penalty relative to
$\hh\to 2b$ or $\ha\to 2b$ tagging
of $\eps_{tt/bb}\equiv [\br(t\to 2jb)]^2\hat\eps\sim 0.25$ (for
$\hat\eps=0.55$), as discussed earlier;
\footnote{These details for 
the $t\anti t$ final state are only 
relevant for $\breff$ defined in Eq.~(\ref{breffdef}) and
the ratios of Eqs.~(\ref{hhhasusy}) and (\ref{hhhattbb})
below, and then only when $\br(\hh,\ha\to t\anti t)$ are relatively large.}
\item $\hh\ha$ with $\hh\to\hl\hl\to 4b$;
\item $\hp\hm$ with $\hp\to t b\to W2b\to 2j2b,$
or the reverse --- $tb$ mass reconstruction will be necessary.
\end{itemize}
In the case of $\hh\ha$ pair production, in determining that the
second (non-tagged) member of the pair decays to $t\anti t$,
we will again demand full $t\anti t$ reconstruction, and we will apply
the extra $\eps_{tt/bb}$ efficiency penalty
relative to $b\anti b$ decay.  This might be somewhat too conservative
an approach, but does simplify our analysis since the event
rates of interest involving $t\anti t+b\anti b$ 
decays will then be proportional to the effective
branching fractions
\begin{equation}
\breff(\hh,\ha\to b\anti b+t\anti t)\equiv \br(\hh,\ha\to b\anti
b)+\eps_{tt/bb}\br(\hh,\ha\to t\anti t)\,.
\label{breffdef}
\end{equation}

Because $\mha\sim \mhh$ over much of parameter space,
we will presume that it is not possible to separate the $\ha$ and $\hh$
from one another. We also stick to our simplifying assumption that
the overall efficiency, $\eps$, associated with detector
coverage, $b$-tagging and so forth does not depend upon
the final state, except that in the case of $t\anti t$ decay
we include an extra $\hat \eps$ in $\eps_{tt/bb}$,
as discussed above and as incorporated through $\breff$ defined in
Eq.~(\ref{breffdef}). With these assumptions, the following
ratios of branching fractions can be extracted directly from
experimental observations using the measured
values of $\br(\hl\to b\anti b)$ and $\br(t\to 2j b)$.
\begin{eqnarray}
&{\br(\hh\to {\rm SUSY})\breff(\ha\to b\anti b+t\anti t)+\br(\ha\to 
{\rm SUSY})\breff(\hh\to
b\anti b+t\anti t) \over \breff(\hh\to b\anti b+t\anti t) 
\breff(\ha\to b\anti b+t\anti t)}
&\label{hhhasusy} \\
&{\br(\hh\to t\anti t )\br(\ha\to b\anti b)+\br(\ha\to t\anti t)\br(\hh\to
b\anti b) \over \br(\hh\to b\anti b) \br(\ha\to b\anti b)}
&\label{hhhattbb} \\
&{\br(\hh\to \hl\hl)\br(\ha\to b\anti b)\over \br(\hh\to b\anti b)
\br(\ha\to b\anti b)}
&\label{hhhlhl}\\
&{\br(\ha\to Z\hl)\br(\hh\to b\anti b)\over \br(\hh\to b\anti b)
\br(\ha\to b\anti b)}
&\label{hazhl}\\
&{\br(\hp \to {\rm SUSY})\br(\hm\to b\anti t)+\br(\hm\to {\rm SUSY})\br(\hp\to
t\anti b)\over  \br(\hp\to t\anti b)\br(\hm\to b\anti t)}
&\label{hphmsusy}\\
&{\br(\hp \to {\taup \nu})\br(\hm\to b\anti t)+\br(\hm\to {\taum \nu})\br(\hp\to
t\anti b)\over  \br(\hp\to t\anti b)\br(\hm\to b\anti t)}
&\label{hptaunu}\\
&{\br(\hp \to {\hl\wp})\br(\hm\to b\anti t)+\br(\hm\to {\hl\wm})\br(\hp\to
t\anti b)\over  \br(\hp\to t\anti b)\br(\hm\to b\anti t)}
&\label{hpwhl}
\end{eqnarray}
As a shorthand, we will employ the notations
\begin{equation}
\left\langle{2\br(\hh ,\ha\rta SUSY)\over
\breff(\hh ,\ha\rta b \anti b,t\anti t)}\right\rangle\,,~~~
\left\langle{2\br(\hh ,\ha\rta t\anti t)\over
\br(\hh ,\ha\rta b \anti b)}\right\rangle\,,
\label{sh1}
\end{equation}
for the ratios of Eqs.~(\ref{hhhasusy})
and (\ref{hhhattbb}), respectively.  The ratios of
Eqs.~(\ref{hphmsusy})-(\ref{hpwhl}) reduce to 
\begin{equation}
{2\br(\hp\to {\rm SUSY},\tau^+\nu,\wp\hl)\over\br(\hp\to t\anti b)}\,,
\label{sh2}
\end{equation}
respectively. We retain both $b\anti b$ and $t\anti t$ final states in
Eq.~(\ref{hhhasusy}), using the combination defined
in $\breff$, in order that we may assess the importance
of SUSY decays both in regions where $b\anti b$ decays of
the $\hh,\ha$ are dominant and in regions where $t\anti t$ decays
are important. 

In estimating the accuracy with which these ratios can be
measured experimentally, it is important to keep track
of the actual final state in which the observation occurs
and the effective efficiency for observing that final state.
We make this explicit below.
\begin{itemize}
\item
The event rate for the numerator of Eq.~(\ref{hhhasusy}) 
is obtained by multiplying the rate for $\hh\ha$ pair
production by $\eps$ (the overall efficiency factor) times
the indicated sum of branching ratio products:
$[\br(\hh\to {\rm SUSY})\breff(\ha\to b\anti b+t\anti t)+\br(\ha\to 
{\rm SUSY})\breff(\hh\to b\anti b+t\anti t)]$.
\item
The numerator of Eq.~(\ref{hhhlhl})
must be measured in the final state in which both $\hl$'s decay
to $b\anti b$.  Thus, the event rate associated with
determining the numerator is obtained by
multiplying the $\hh\ha$ pair production rate by
a factor of $[\br(\hl\to b\anti b)]^2$
times the overall efficiency $\eps$ times
$\br(\hh\to \hl\hl)\br(\ha\to b\anti b)$. 
\item
The event rate associated
with measuring the numerator of Eq.~(\ref{hhhattbb}) is obtained using
a factor of $\eps\eps_{tt/bb}=\eps\hat\eps [\br(t\to 2jb)]^2$
times $\br(\hh\to t\anti t)\br(\ha\to b\anti b)+\br(\ha\to t\anti
t)\br(\hh\to b\anti b)$. 
\item
The event rate for the numerator of Eq.~(\ref{hazhl}) is computed
using the factor 
$\eps\br(\hl\to b\anti b)\br(\ha\to Z\hl)\br(\hh\to b\anti b)$.
This implicitly assumes that we can sum
over all $Z$ decays, as would be possible since
the $Z$ mass can be reconstructed from the c.m. $\rts$ value and the 
momenta of the four $b$'s.
\item
The event rate for the numerator of Eq.~(\ref{hphmsusy})
is obtained by multiplying the $\hp\hm$ event rate
by the factor $\eps\br(t\to 2j b)[\br(\hp\to {\rm SUSY})\br(\hm\to b\anti t)+
\br(\hm\to {\rm SUSY})\br(\hp\to t\anti b)]$.
\item
The event rate for the numerator of Eq.~(\ref{hptaunu}) is computed
by multiplying the pair rate by
$\eps\br(t\to 2j b)
[\br(\hp \to {\taup \nu})\br(\hm\to b\anti t)+\br(\hm\to {\taum \nu})\br(\hp\to
t\anti b)]$.
\item
The rate for the numerator of Eq.~(\ref{hpwhl}) is computed using
the factor
$\eps\br(\hl\to b\anti b)\br(t\to 2jb)
[\br(\hp\to\hl\wp)\br(\hm\to b\anti t)+
\br(\hm\to \hl\wm)\br(\hp\to t\anti b)]$.
\item
The factors for the denominators are obtained by multiplying
the indicated branching ratio product by $\eps$ in the case
of the neutral Higgs ratios, and by $\eps[\br(t\to 2j b)]^2$
in the case of charged Higgs ratios.
\item
When dividing the SUSY collection of final states (as simply
identified by missing energy) into subcategories of a certain
number of leptons and/or jets, the full set of appropriate
branching ratios are included in all the chain decays leading
to the specified final state.
\end{itemize}
As noted, the overall factor of $\eps$ common
to all rates is incorporated by reducing the full luminosity to the
effective luminosity $L_{\rm eff}$.  Rates for 
the standard $L_{\rm eff}=80\fbi$ are thus obtained by computing
the pair production cross section, multiplying by $L_{\rm eff}$
and then including all the above factors after removing the
overall multiplicative $\eps$ contained in each.
The bottom line is that
even though we plot the ratios listed, 
the statistical errors we shall discuss
will be based on the actual number of events as obtained
according to the above-outlined procedures.

The utility of the above ratios derives from the following general features.
The 1st ratio is primarily a function of $\tanb$. The
2nd provides an almost direct determination of $\tanb$
since $t\anti t/b\anti b$ is roughly proportional to $\cot^4\beta$
in the MSSM. 
The ratios of Eqs.~(\ref{hphmsusy}) 
and (\ref{hptaunu}) both exhibit substantial and rather orthogonal variation
as a function of $\tanb$ and $\mhalf$.
The ratio of Eq.~(\ref{hhhlhl}) is proportional to
the relative strength of the $\hh\to\hl\hl$ trilinear coupling as compared
to the $\hh\to b\anti b$ coupling.  This could be the first direct probe
of Higgs trilinear couplings.
The ratios of Eqs.~(\ref{hazhl}) and (\ref{hpwhl})
would probe the very interesting Higgs--Higgs--vector-boson couplings.
These features will be illustrated shortly.

\subsection{Resolving Ambiguities in Identifying Different Final States}

\indent\indent
Since all SUSY final states will contain substantial missing energy,
the ambiguities in separating SUSY decays from others
are limited. We discuss below the procedures for removing
the only ambiguities that appear to be of importance.

\medskip
\noindent (A)
A potential ambiguity arises in $\ell^++\etmiss$ final states of
the $\hp$ to which the SUSY $\slep^+\snu$ and $\chitil^+\chitil^0$
decay modes and the SM $\hp\to \tau^+\nu\to \ell^+ 3\nu$ decay modes
all contribute. 
The $\hp\to\tau^+\nu_\tau\to \ell^+3\nu$ decay can be identified
using kinematic constraints.  Consider the c.m. system of the decaying
$\hp$ (as determined using incoming beam information and the 
tagged $\hm$ four-momentum). To the extent that $m_\tau$
can be neglected and, therefore, 
the $\tau$ decays collinearly to $\ell^+2\nu$,
all of which move opposite the primary $\nu_\tau$, one must
have $E=|\etmiss|$, where $E$ is the energy of
the observed $\ell$.  SUSY events of any type will normally violate
this constraint. In what follows,
the $\slep^+\snu$ and $\chitil^+\chitil^0$ decays are both included in
the overall SUSY decay rate of the $\hp$. 
\medskip

\noindent (B) In $\hh$ or $\ha$ decay, $\tau^+\tau^-$ decays contribute to the
same $\ell^+\ell^-+\etmiss$ final states to which the SUSY modes
$\slep^+\slep^-$ and $\chitil^+\chitil^-$ contribute. The procedure
for eliminating the $\tau^+\tau^-$ decay is analogous to that
discussed in (A) for removing $\hp\to\tau^+\nu_\tau$ decays.
We again note that, for most events, the $\tau$ mass can
be neglected relative to its momentum.  In the (known)
rest frame of the Higgs, the collinear approximation implies that
the $\ell^+$ and $\ell^-$ and their associated neutrinos travel
in essentially the same directions as the parent $\tau^+$ and $\tau^-$,
respectively. As a result, such events must have $|E_+-E_-|=|\etmiss|$,
where $E_{\pm}$ are the observed energies of the $\ell^{\pm}$
in the Higgs rest frame.  The very non-collinear
SUSY modes would generally be far from approximately satisfying this
constraint.  
Kinematic constraints do not allow an event-by-event separation
of the two SUSY modes, $\slep^+\slep^-$ and $\chitil^+\chitil^-$,
in the $\ell^+\ell^-+\etmiss$ final state. These are
lumped together as part of the overall SUSY decay branching fraction.
\medskip

\noindent (C) Events in which the unreconstructed Higgs boson/decay
is $\hh~{\rm or}\ha\to t\anti t\to  \ell\nu 2j 2b$
or $\hpm\to t b\to \ell \nu b\anti b$  can be eliminated by
using the incoming beam energy/momentum 4-vector, subtracting
the momenta of all visible final state leptons and jets, and computing
the invariant mass of the resulting difference 4-vector.  This would belong
to the $\nu$ in the above cases. A cut requiring
a substantial value would eliminate the above final states
and be highly efficient in retaining true SUSY decays.
For parameters such that
the rates for single neutrino events (as defined by the above
procedure and requiring a small value for the difference 4-vector
mass) are significant, we shall find that
the $\hh,\ha\to t\anti t$ and $\hp\to t\anti b$
branching fractions can be directly measured with reasonable accuracy
(using all-jet modes). The predicted single neutrino rate could
then be compared to that observed as a further check.
Events where the unreconstructed Higgs/decay 
is $\hh~{\rm or}~\ha\to t\anti t\to  2\ell 2\nu 2b$  cannot be eliminated
by the above technique. However, the branching fractions,
$\br(\hh,\ha\to t\anti t)$, measured in all-jet final
states can be employed to make
an appropriate correction. The single neutrino rates, as defined
above, may allow a double-check of the all-jet final state determinations
of the $t\anti t$ branching fractions.

\medskip
\noindent (D) Other ambiguities include events in which
the second Higgs boson/decay is: $\ha\to Z\hl\to Z\tauptaum$, where the
$Z$ decays invisibly or to $\tauptaum$; $\hpm\to \wpm\hl\to \wpm \tauptaum$
where the $\wpm$ decays leptonically; and $\hh\to \hl\hl\to\tauptaum\tauptaum$.
The common characteristic of all these is the presence of missing energy 
from $\tau$, $W$ and/or $Z$ decays that is due to more than a single neutrino
and that makes it impossible to either directly or indirectly reconstruct
the mass of the $\hl$, $W$ and/or $Z$.
However, the event rates for these processes are so low that they 
can be included in SUSY decays without any visible alteration of the effective
SUSY branching fraction. Further, whenever the $\hh\to \hl\hl$,
$\ha\to Z\hl$ or $\hp\to\wpm\hl$ decays are significant, we shall
see that at least a rough measurement of the corresponding branching
ratio will be possible in all-jet modes.  Given the known
$Z\to \nu\anti \nu$ and $\hl\to \tauptaum$ branching fractions, a
correction could then be made using a Monte Carlo simulation.

\subsection{Ratio Contours, Error Estimates and Model Discrimination}

\indent\indent
In order to determine how well we can measure
the ratios of Eqs.~(\ref{hhhasusy}), (\ref{hhhattbb}), 
(\ref{hhhlhl}), (\ref{hazhl}),
(\ref{hphmsusy}), (\ref{hptaunu}), and (\ref{hpwhl}),
we have proceeded as follows. 
For each of the six scenarios (\DM, \DP, \ldots)
and for a given $(\mhalf,\tanb)$ choice within the allowed parameter
space of a given scenario, we first compute the expected 
number of events available for determining the numerator or
denominator of each ratio. The ingredients (such as branching
ratios and efficiencies) in the event number computations for each channel
were given earlier.
The expected number of events in the numerator
or denominator is taken as the 
mean value in determining a Poisson distribution 
for that event number; if the mean number of events is $\geq
30$, then we use a Gaussian approximation to the distribution. 
From the event number distributions we compute the probability 
for the numerator and denominator of each ratio to take on given values.
(We fluctuate the event numbers and then correct for branching fractions
and efficiencies.)
The probability of the resulting value for the ratio is then simply
the product of these probabilities. The probabilities for different
combinations that yield the same value for the ratio are summed.
In this way, we obtain a probability for every possible value of the ratio.
These probabilities are re-ordered so as to form a distribution.
The lower (upper) limit for the ratio at this $(\mhalf,\tanb)$ value is
then found by adding up the probabilities, starting from zero, until the sum of
is 15.9\% (84.1\%). In other words, the confidence level that the
true value of the ratio is higher (lower) than the lower (upper) limit
is 84.1\%. These would be the $\pm1\sigma$ upper/lower limits for the
ratio in the limit where the distribution of the ratio is normal.

In computing the number of events available for determining the
numerator or denominator (or one of the independent contributions thereto)
we include only fully reconstructable final states
for the tagged Higgs boson. The branching ratios and efficiency factors
were detailed below Eq.~(\ref{sh2}). We presume $\leff=80\fbi$
at $1\tev$ ($\leff=400\fbi$ at $4\tev$). The efficiency factor, $\eps$,
included in $\leff$ should reflect efficiencies
associated with identifying a particular type of event
in such a way as to eliminate backgrounds, \eg\ via $b$-tagging,
cuts on $\etmiss$, and so forth; the $\epsilon$ appropriate
to the current situation where one of the Higgs must be clearly
`tagged' (as defined earlier)
will probably be smaller than that appropriate to simply
discovering a signal, given the need to clearly separate
different types of final states from one another.
Thus, the above $\leff=80\fbi$ ($400\fbi$) values probably would
only be achieved after several years of running. 
We re-emphasize that an implicit approximation to our approach is that 
$\leff$ is the same for all the observationally/statistically
independent final states that appear in 
the numerator and denominator of a given ratio.
[Aside from our special $\hat\eps=0.55$ correction for $t\anti t$
reconstruction, the only explicitly channel-dependent factors that
have been included are the relevant branching fractions, as detailed 
below Eq.~(\ref{sh2}).]
Presumably, this will not be true in practice, but it is 
at least a reasonable first approximation.  Full detector
specification and simulation would be necessary to do better.

\begin{figure}[[htbp]
\vskip 1in
\let\normalsize=\captsize   
\begin{center}
\centerline{\psfig{file=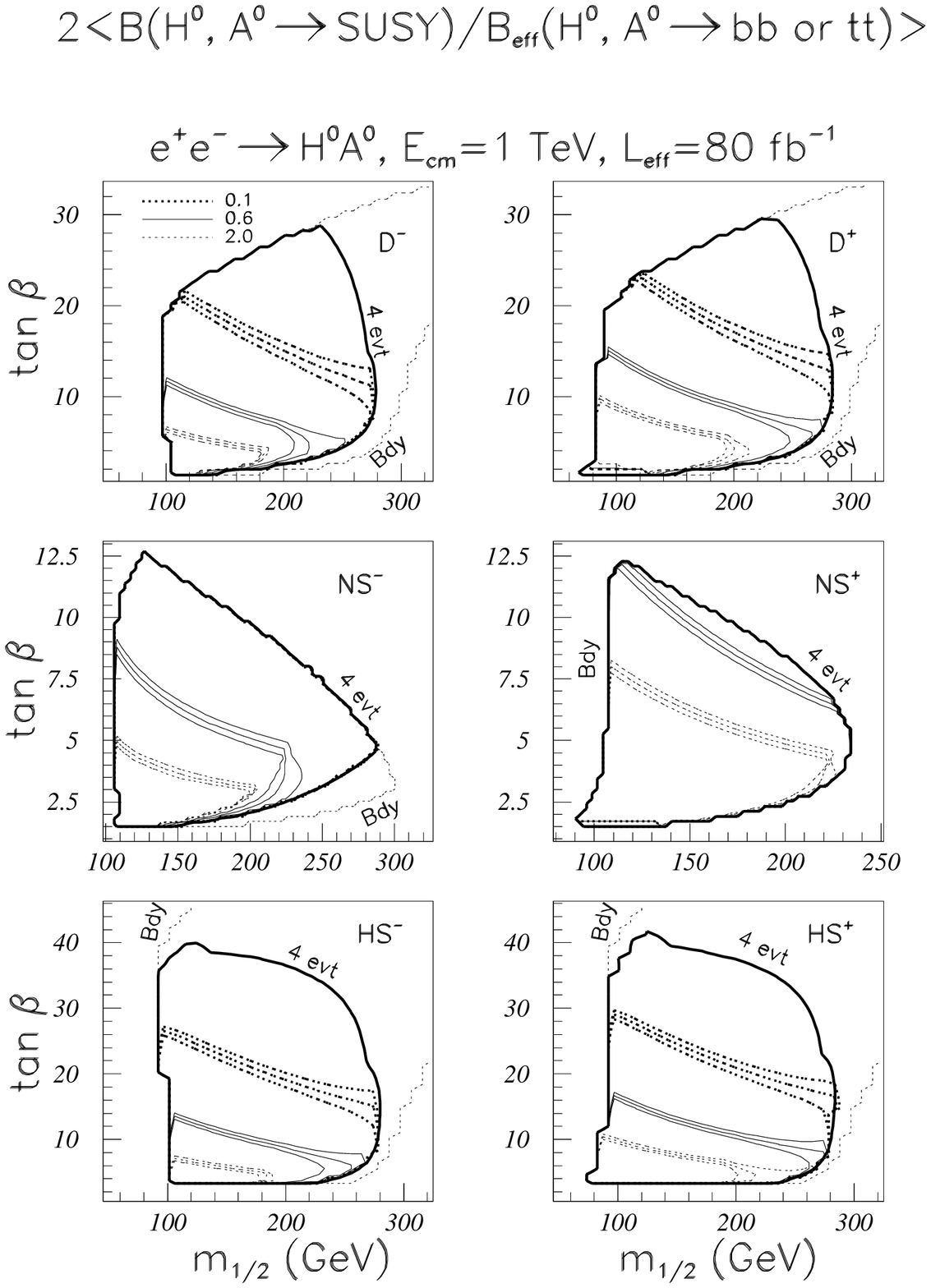,width=10.5cm}}
\smallskip
\begin{minipage}{12.5cm}       
\caption{
We plot contours, along which the ratio of Eq.~(\ref{hhhasusy})
has a given constant value,
within the constraint/kinematically allowed
$(\mhalf,\tanb)$ parameter space of the \DM, \DP, \NSM, \NSP, \HSM, and
\HSP\ models. Results are shown for the same three central values
for all models. For each central value, three lines are drawn. The central
line is for the central value. The other two lines 
are contours for which the ratio deviates by $\pm1\sigma$ statistical
error from the central value.
Bold lines indicate the boundary beyond which fewer
than 4 events are found in the final states used to measure the numerator
of the ratio.
}
\label{fhhhasusy}
\end{minipage}
\end{center}
\end{figure}

\begin{figure}[[htbp]
\vskip 1in
\let\normalsize=\captsize   
\begin{center}
\centerline{\psfig{file=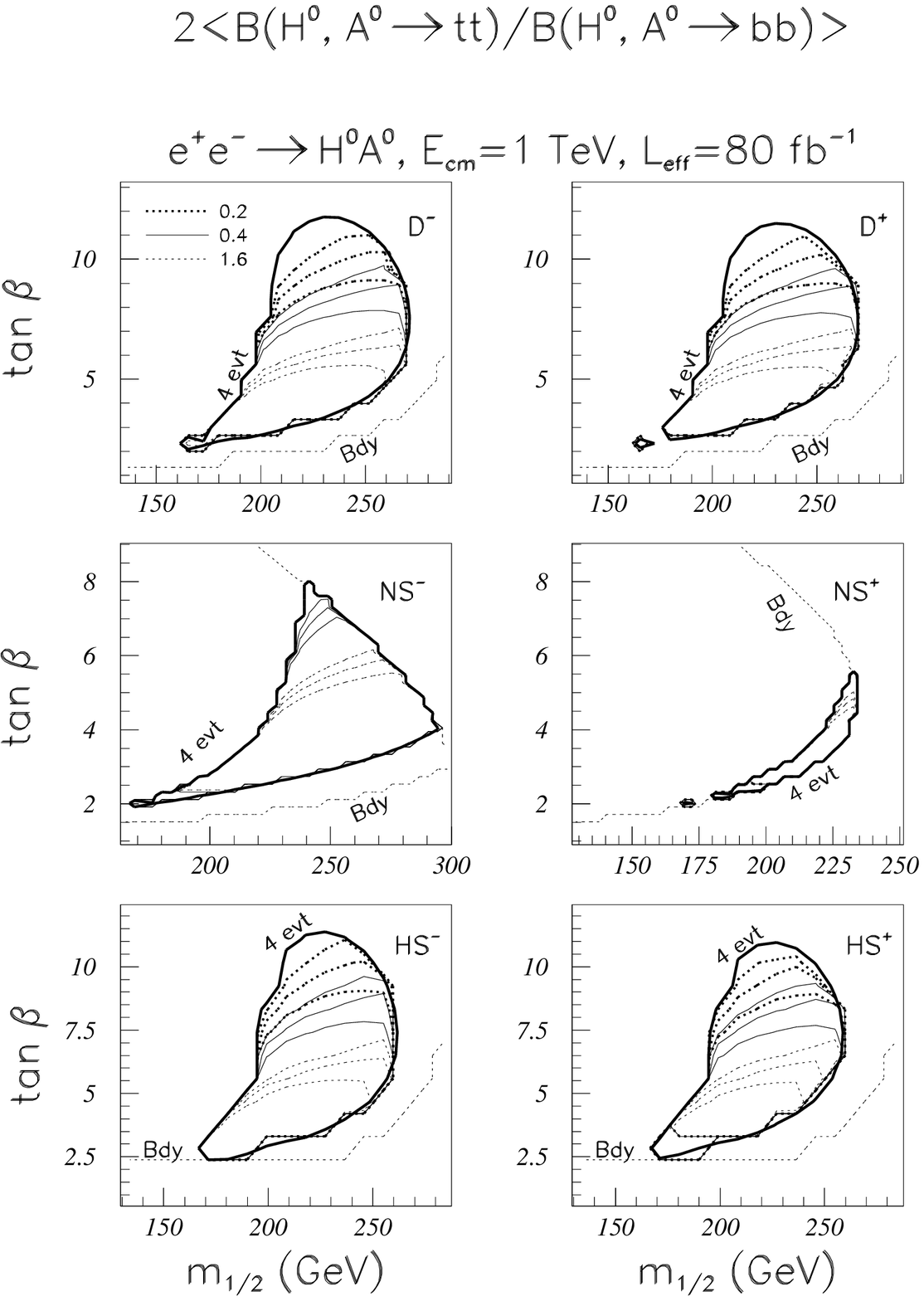,width=10.5cm}}
\smallskip
\begin{minipage}{12.5cm}       
\caption{
As in Fig.~\ref{fhhhasusy}, but for the ratio of Eq.~(\ref{hhhattbb}).
}
\label{fhhhattbb}
\end{minipage}
\end{center}
\end{figure}

\begin{figure}[[htbp]
\vskip 1in
\let\normalsize=\captsize   
\begin{center}
\centerline{\psfig{file=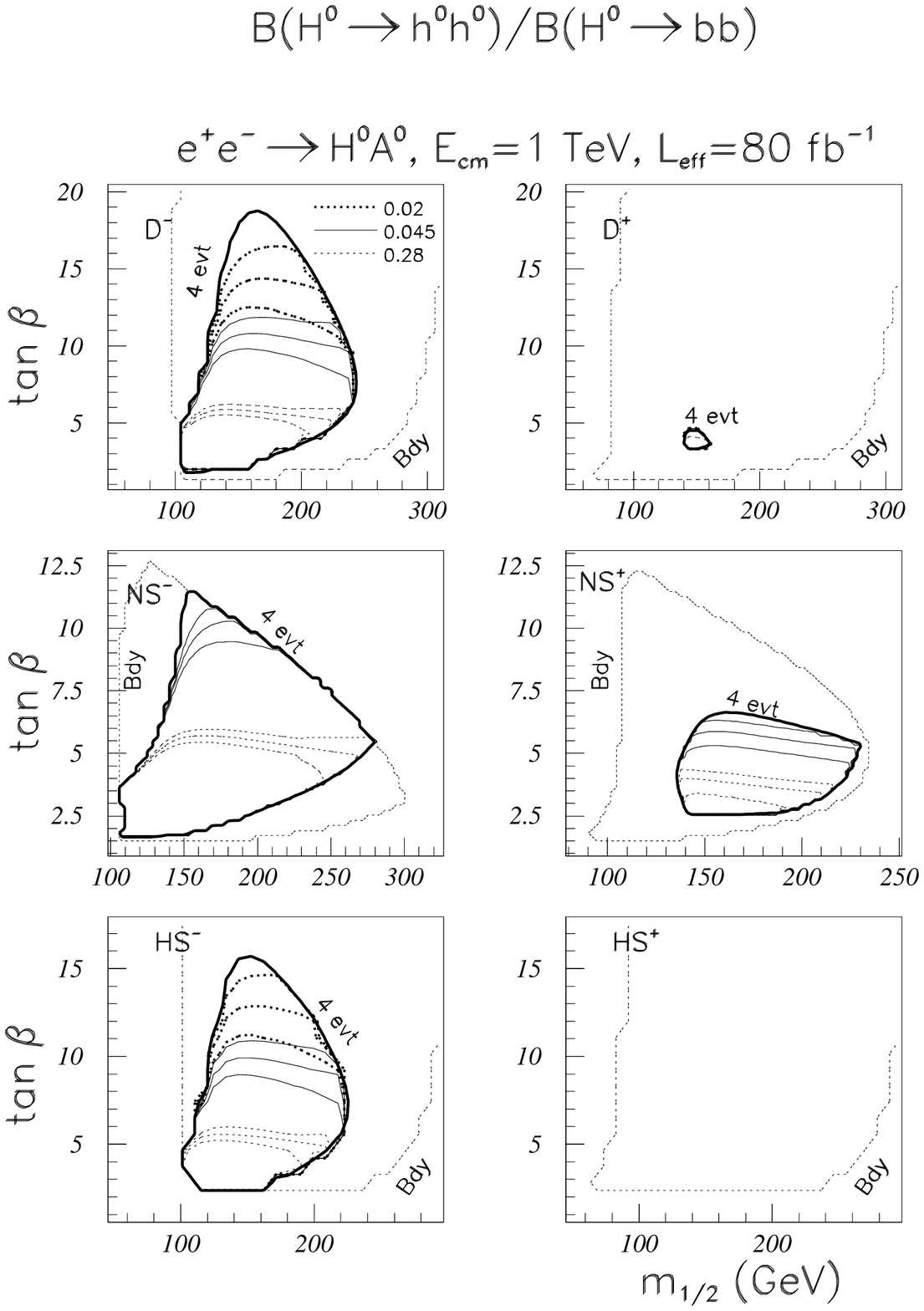,width=10.5cm}}
\smallskip
\begin{minipage}{12.5cm}       
\caption{
As in Fig.~\ref{fhhhasusy}, but for the ratio of Eq.~(\ref{hhhlhl}).
}
\label{fhhhlhl}
\end{minipage}
\end{center}
\end{figure}

\begin{figure}[[htbp]
\vskip 1in
\let\normalsize=\captsize   
\begin{center}
\centerline{\psfig{file=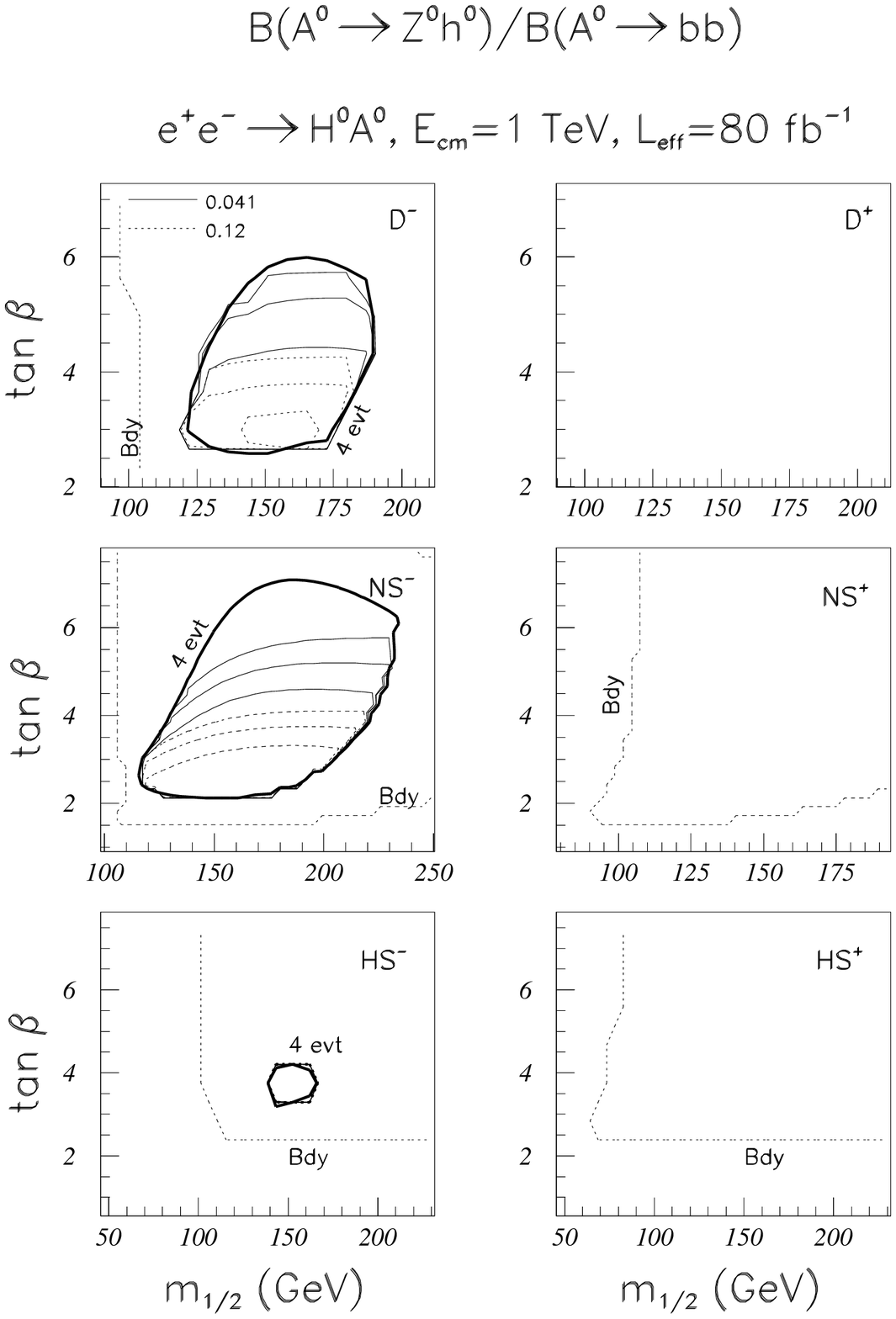,width=10.5cm}}
\smallskip
\begin{minipage}{12.5cm}       
\caption{
As in Fig.~\ref{fhhhasusy}, but for the ratio of Eq.~(\ref{hazhl}).
}
\label{fhazhl}
\end{minipage}
\end{center}
\end{figure}

\begin{figure}[[htbp]
\vskip 1in
\let\normalsize=\captsize   
\begin{center}
\centerline{\psfig{file=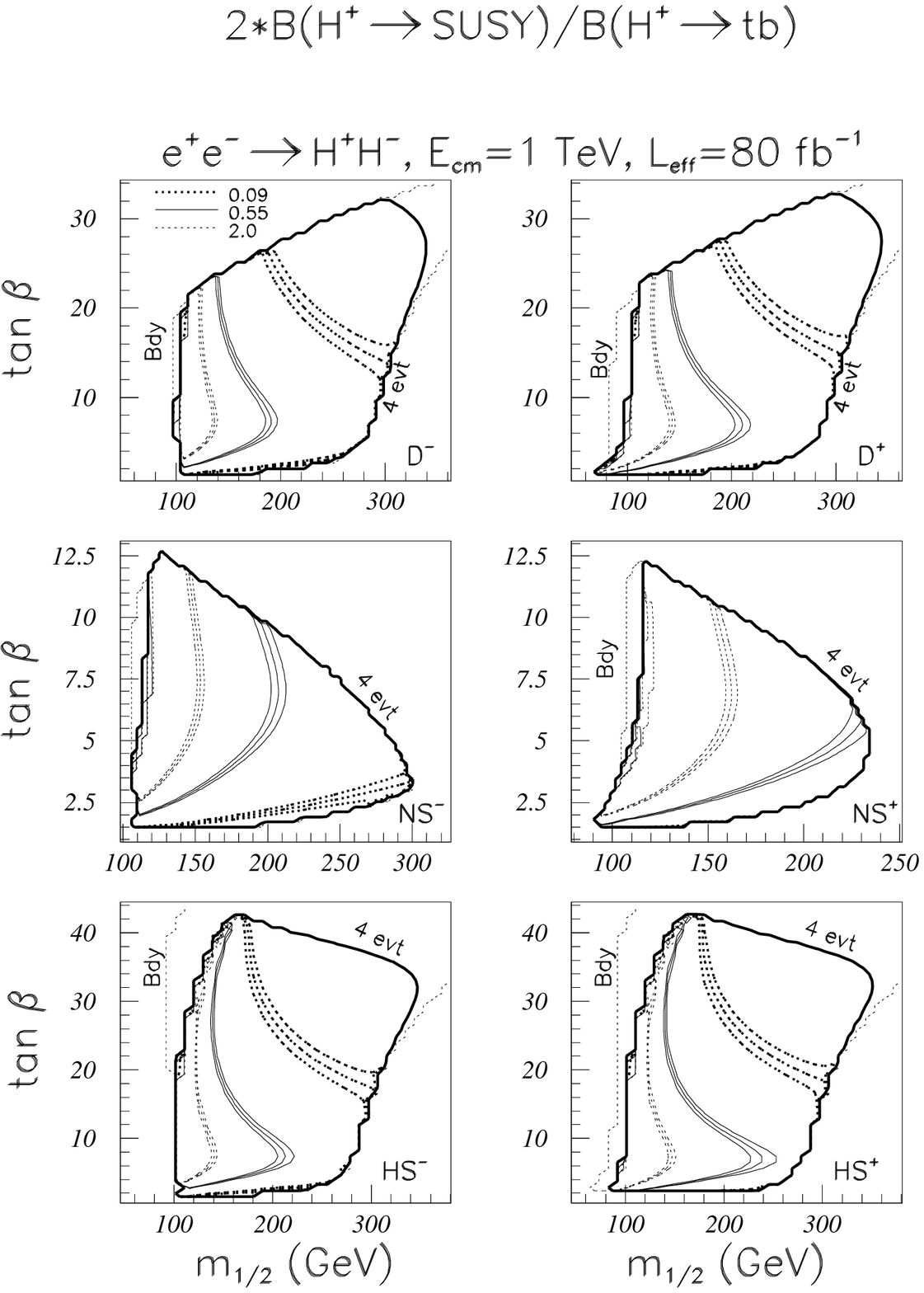,width=10.5cm}}
\smallskip
\begin{minipage}{12.5cm}       
\caption{
As in Fig.~\ref{fhhhasusy}, but for the ratio of Eq.~(\ref{hphmsusy}).
}
\label{fhphmsusy}
\end{minipage}
\end{center}
\end{figure}

\begin{figure}[[htbp]
\vskip 1in
\let\normalsize=\captsize   
\begin{center}
\centerline{\psfig{file=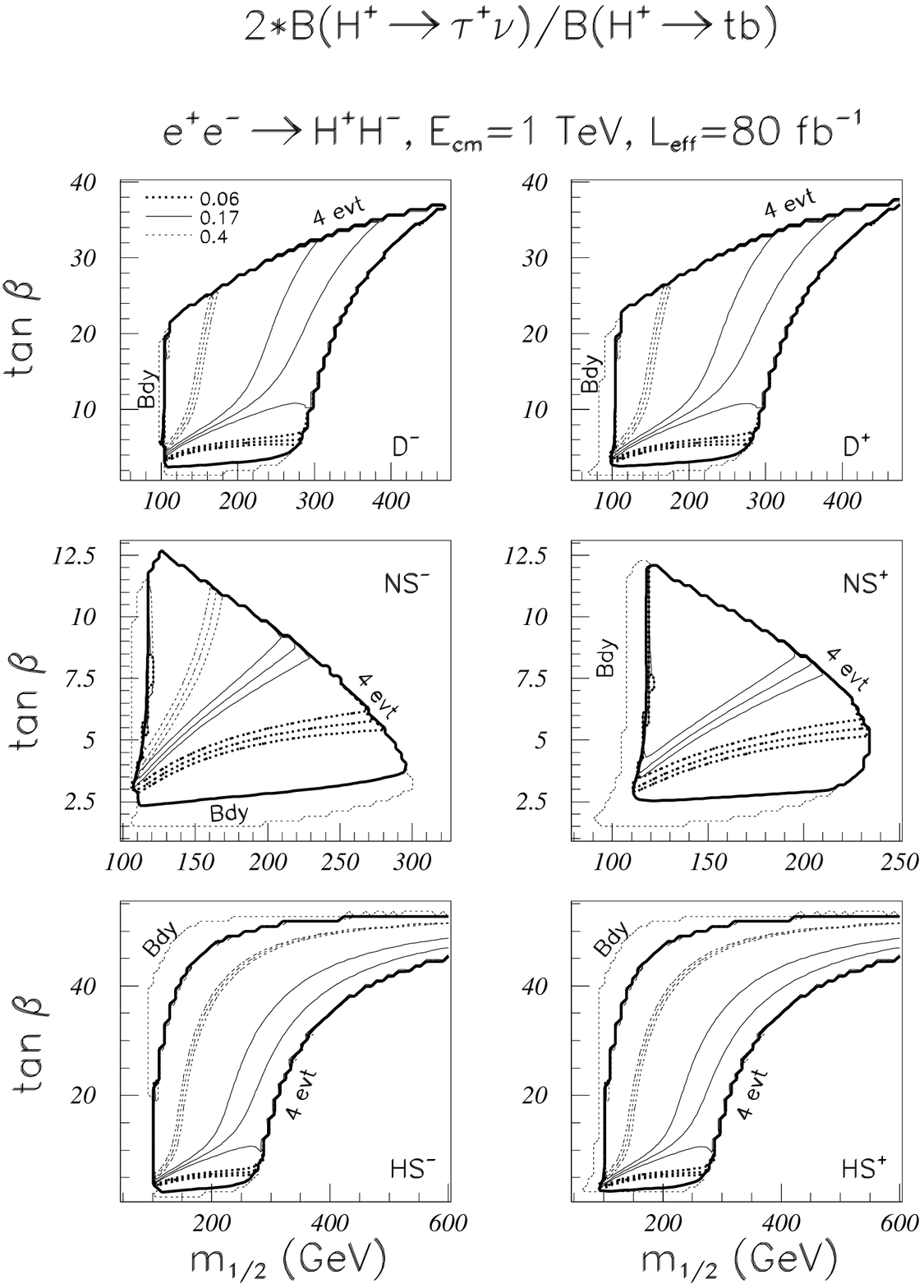,width=10.5cm}}
\smallskip
\begin{minipage}{12.5cm}       
\caption{
As in Fig.~\ref{fhhhasusy}, but for the ratio of Eq.~(\ref{hptaunu}).
}
\label{fhptaunu}
\end{minipage}
\end{center}
\end{figure}

\begin{figure}[[htbp]
\vskip 1in
\let\normalsize=\captsize   
\begin{center}
\centerline{\psfig{file=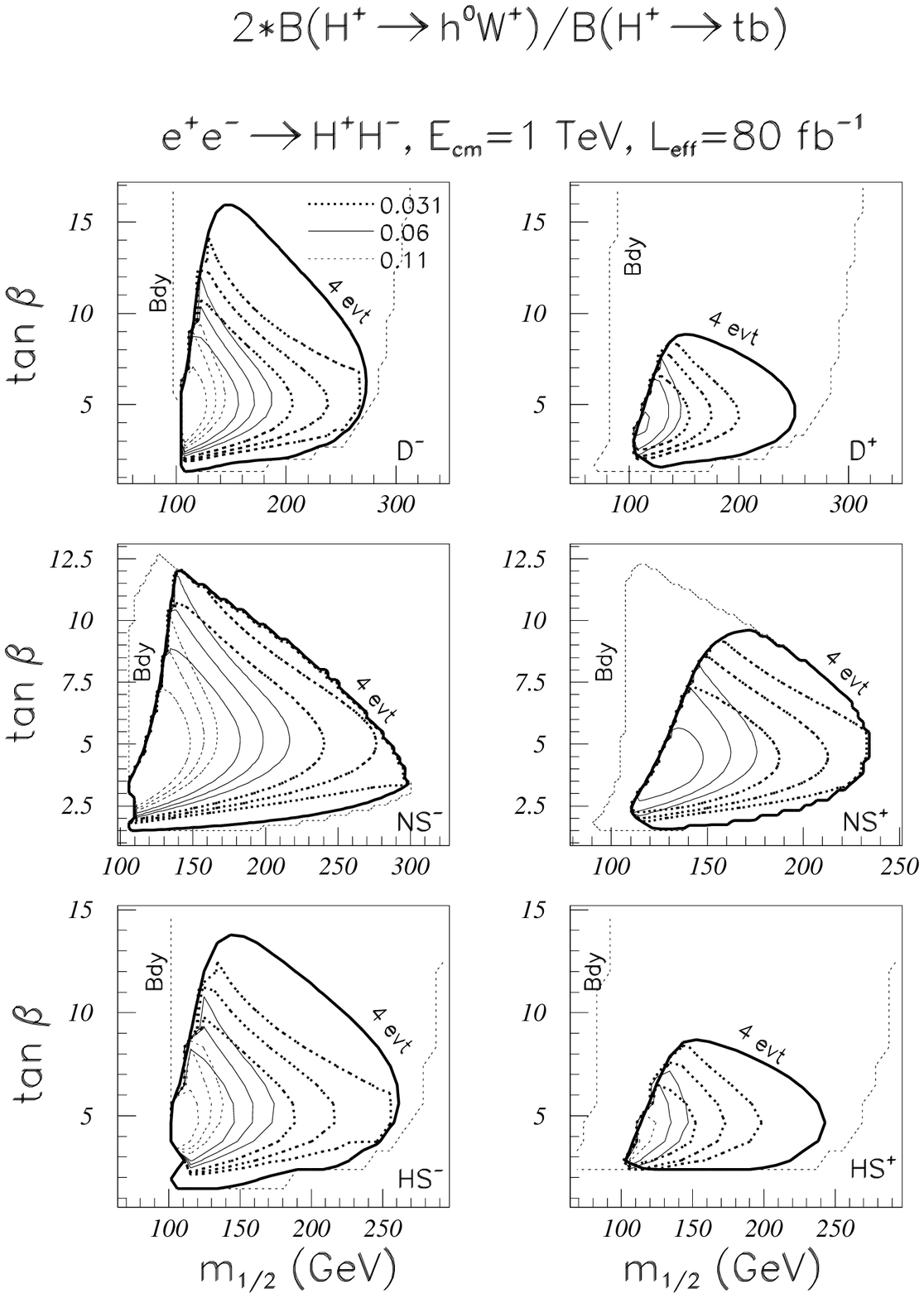,width=10.5cm}}
\smallskip
\begin{minipage}{12.5cm}       
\caption{
As in Fig.~\ref{fhhhasusy}, but for the ratio of Eq.~(\ref{hpwhl}).
}
\label{fhpwhl}
\end{minipage}
\end{center}
\end{figure}

In Figs.~\ref{fhhhasusy}, \ref{fhhhattbb}, \ref{fhhhlhl}, \ref{fhazhl},
\ref{fhphmsusy}, \ref{fhptaunu}, and \ref{fhpwhl}                    
we plot contours of constant values for
the ratios of Eqs.~(\ref{hhhasusy}), (\ref{hhhattbb}), 
(\ref{hhhlhl}), (\ref{hazhl}),
(\ref{hphmsusy}), (\ref{hptaunu}), and (\ref{hpwhl})
within the $\rts=1\tev$ constraint/kinematically
allowed $(\mhalf,\tanb)$ parameter space.  Associated
with each such contour, we give two additional contours
showing how much the $\tanb$ value at a given (known) value of $\mhalf$
would have to change in order to reproduce the values
obtained for deviations in the ratio 
at the $\pm1\sigma$ statistical level. [As previously explained,
$\pm1\sigma$ is our short hand phrase for deviations such that
the ratio has 84.1\% probability of being lower (higher) than
the upper (lower) limit.]
We do not consider errors when there are fewer
than 4 events that can be used to determine
the numerator for one of these ratios.
The 4-event contours are indicated on the figures.

Consider first the relative SUSY branching
ratio contours of Eqs.~(\ref{hhhasusy}) and (\ref{hphmsusy})
displayed in Figs.~\ref{fhhhasusy} and \ref{fhphmsusy}, respectively.
For most points in parameter space, a simultaneous measurement
of the two ratios will determine a fairly small and 
unique region in the parameter space of a given
model that is simultaneously consistent with both measurements
at the $1\sigma$ level.

\begin{figure}[[htbp]
\vskip 1in
\let\normalsize=\captsize   
\begin{center}
\centerline{\psfig{file=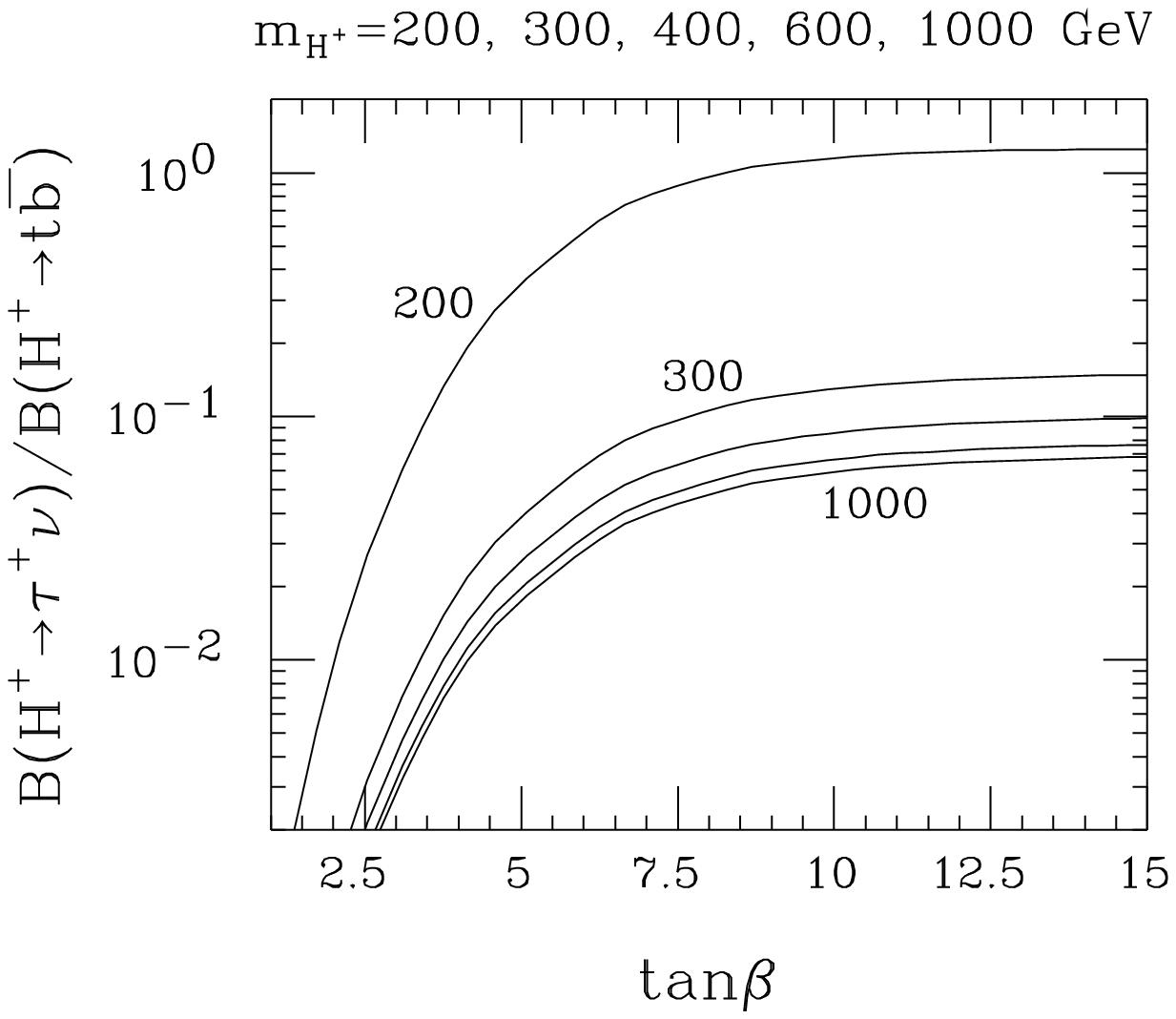,width=12.5cm}}
\smallskip
\begin{minipage}{12.5cm}       
\caption{
The ratio $\br(\hp\to \taup\nu)/\br(\hp\to t\anti b)$ 
computed at tree level for $\mt=175\gev$ and  $\mb=4\gev$ as a function
of $\tanb$ for $\mhp=200$, 300, 400, 600, and 1000 GeV.
}
\label{taunutbratio}
\end{minipage}
\end{center}
\end{figure}

\begin{figure}[[htbp]
\vskip 1in
\let\normalsize=\captsize   
\begin{center}
\centerline{\psfig{file=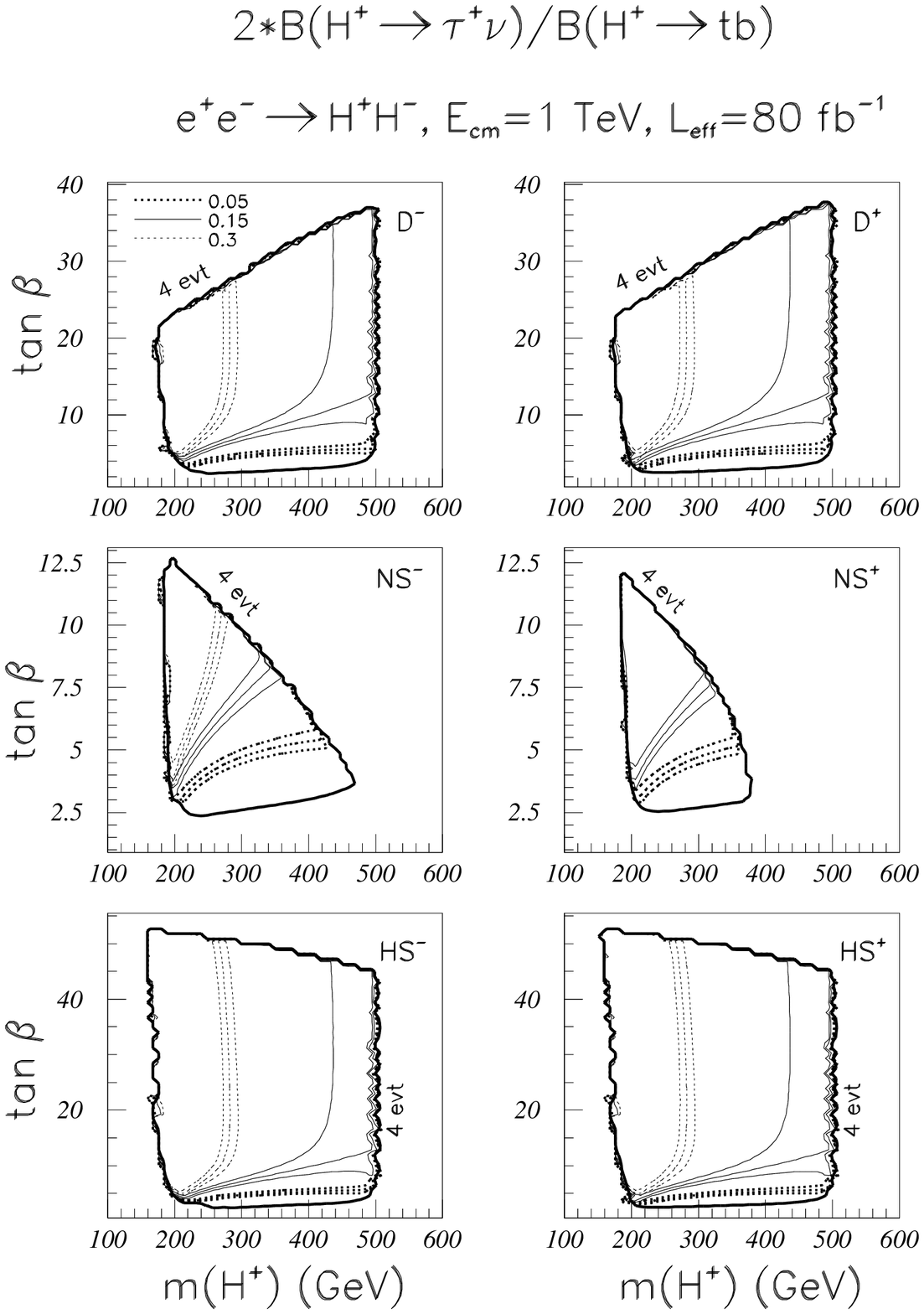,width=12.5cm}}
\smallskip
\begin{minipage}{12.5cm}       
\caption{
Contours of the ratio $\br(\hp\to \taup\nu)/\br(\hp\to t\anti b)$ and
its associated $\pm 1\sigma$ contours are plotted as a function of $\tanb$
and $\mhp$ for $\leff=80\fbi$ at $1\tev$.
}
\label{taunuvshp}
\end{minipage}
\end{center}
\end{figure}

\begin{figure}[[htbp]
\vskip 1in
\let\normalsize=\captsize   
\begin{center}
\centerline{\psfig{file=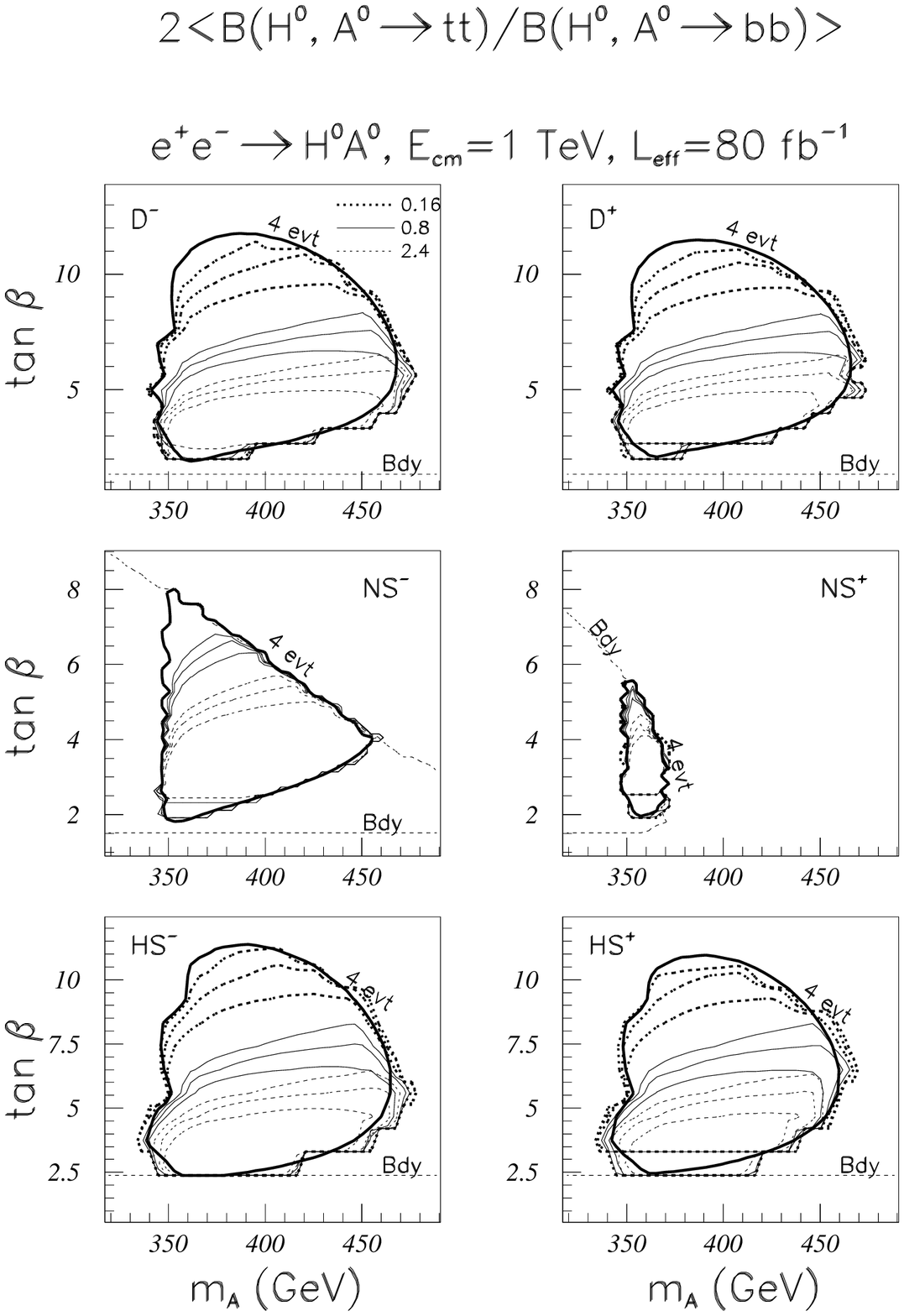,width=12.5cm}}
\smallskip
\begin{minipage}{12.5cm}       
\caption{
Contour of the ratio 
$2\langle\br(\hh ,\ha\to t\anti t)/\br(\hh, \ha\to b\anti b)\rangle$ 
its associated $\pm 1\sigma$ contours are plotted as a function of $\tanb$
and $\mha$ for $\leff=80\fbi$ at $1\tev$.
}
\label{ttvsha}
\end{minipage}
\end{center}
\end{figure}

If $\tanb$ is not large, then measuring
$\br(\hp\to \taup\nu)/\br(\hp\to t\anti b)$ via
the ratio of Eq.~(\ref{hptaunu}) can provide a second determination
of $\tanb$.  The dependence of this ratio on $\tanb$ for a
selection of $\mhp$ values is illustrated in Fig.~\ref{taunutbratio}.
There, the ratio is computed at tree level.  We see that
the ratio depends sensitively on $\tanb$ at fixed $\mhp$ for $\tanb\lsim 6$.
For such $\tanb$ values, measurement of the ratio provides an
excellent $\tanb$ determination.
However, when $\tanb$ is large the $\taup\nu/t\anti b$ ratio 
becomes independent of $\tanb$ and sensitivity is lost.
Note also that the ratio becomes independent of $\mhp$ when $\mhp$ is large.
Thus, when both $\mhp$ and $\tanb$ are large, this ratio will provide
little information regarding location in parameter space.

Contours of constant $\br(\hp\to\taup\nu)/\br(\hp\to t\anti b)$ 
in $(\mhalf,\tanb)$ parameter space are displayed in
Fig.~\ref{fhptaunu}. It is also useful to plot these same contours
in $(\mhp,\tanb)$ parameter space, as done in Fig.~\ref{taunuvshp}.
In both figures, one observes a change
from horizontal to vertical contours as one moves from low $\tanb$
and large $\mhalf$ (equivalent to large $\mhp$) to high $\tanb$ and 
small $\mhalf$ (implying small $\mhp$).
The horizontal nature of the contours at 
large $\mhalf,\mhp$ and small $\tanb$ 
can be understood from Fig.~\ref{taunutbratio}.
As already briefly noted, this figure shows that when $\tanb$ is small,
small changes in $\tanb$ yield large changes in the ratio,
whereas there is little sensitivity to changes in $\mhp$ at fixed $\tanb$
when $\mhp$ is large. In contrast, for small $\mhp$ 
Fig.~\ref{taunutbratio} shows that small changes in $\mhp$ 
produce large changes in the ratio,
whereas there is almost no sensitivity to $\tanb$ when $\tanb$ is large.
As a result the contours 
in Figs.~\ref{fhptaunu} and \ref{taunuvshp}
are vertical at small $\mhp$ when $\tanb$ is large.
The wide separation between the central and $\pm 1\sigma$
contours when $\mhalf$ and $\tanb$ are both large is a reflection
of the constancy of this ratio (as displayed
in Fig.~\ref{taunutbratio})  when both $\tanb$ and $\mhp$ are large.
Outside the region where $\tanb$ and $\mhp$ are both large,
the $\taup\nu/t\anti b$ contours are roughly
`orthogonal' to those for the two SUSY ratios discussed earlier.

In general, it is apparent that the contours for the ratios
of Eqs.~(\ref{hhhasusy}), (\ref{hphmsusy}) and (\ref{hptaunu})
in the $(\mhalf,\tanb)$ plane are all oriented rather differently.
This means that, in combination, these three relative Higgs branching fractions 
provide a fairly powerful check of the consistency
of a given model, as well as a very definite determination of the value 
of $\tanb$ that is required for a particular value of $\mhalf$ in the model.
We have already noted that $\mhalf$ will be accurately determined 
in a given model by the neutralino and chargino masses, 
and that the measured $\mha$ will generally provide a $\tanb$ determination.
This determination of $\tanb$ from the masses and the value
for $\tanb$ required for consistency with the above three
ratios of branching fractions are usually not consistent with
one another for an incorrect model choice.

Additional discrimination power between the correct and an incorrect
model choice is possible if we resolve the SUSY rates in
Eqs.~(\ref{hhhasusy}) and (\ref{hphmsusy}) into final states with
a fixed number of leptons plus any number of jets (including 0) 
plus missing energy. Thus, instead of the single ratio of Eq.~(\ref{hhhasusy}),
where SUSY was defined to be the sum over all supersymmetric decay channels,
it will prove useful to
consider the three ratios obtained by dividing SUSY into
the (i) $[0\ell][\geq0j]$, (ii) $[1\ell][\geq0j]$ 
and (iii) $[2\ell][\geq0j]$ channels,
where the $[\geq0j]$ notation indates that states with any
number of jets (including 0) are summed over.
Rates with $[\geq 3\ell][\geq0j]$ are negligible.
Similarly, instead of the ratio of Eq.~(\ref{hphmsusy}) we 
will consider the two
ratios obtained by separating SUSY into the channels (i) or (ii) defined
above.  Rates with $[\geq 2\ell][\geq0j]$ are negligible. 
All SUSY final states will have large missing energy.
The five observable 
SUSY ratios so obtained are not very closely correlated, and thus are 
unlikely to be consistent with one another and with the $\taup\nu/t\anti b$
ratio for any but the correct model choice.

Still more discrimination power can be achieved via the
other branching fraction ratios defined in
Eqs.~(\ref{hhhattbb}), (\ref{hhhlhl}), (\ref{hazhl}), and (\ref{hpwhl}).
For example, we see from Fig.~\ref{fhhhattbb} that the $t\anti t/ b\anti b$
ratio is quite sensitive to $\tanb$.  This is even clearer by displaying
the contours in $(\mha,\tanb)$ space, Fig.~\ref{ttvsha}.
The $\hh\to \hl\hl/\hh\to b\anti b$, 
$\ha\to Z\hl/ \ha\to b\anti b$ and $\hp\to \wp\hl/\hp\to t\anti b$
ratios plotted in Figs.~\ref{fhhhlhl}, \ref{fhazhl} and \ref{fhpwhl},
respectively, are also sensitive to $\tanb$.  However, even more
interesting is their sensitivity to the sign of the $\mu$ parameter.
All three ratios are much smaller for $\mu>0$ than for $\mu<0$
[at a fixed $(\mhalf,\tanb)$ location].
These differences derive almost entirely from a large decrease
in the $\hh\to \hl\hl$, $\ha\to Z\hl$ and $\hp\to \wp\hl$ couplings,
respectively, as the sign of $\mu$ is changed from $+$ to $-$.
(In the case of the $\hh\to\hl\hl$ coupling, this decrease is largely due
to the change of sign of a radiative correction to the vertex associated
with top, bottom, stop and sbottom loops.  In the
$\ha\to Z\hl$ and $\hp\to \wp\hl$ cases, the large decrease is
a tree-level effect.)  Together, these three ratios will provide
significant discrimination between scenarios with the opposite sign of $\mu$.

\subsection{Quantitative Strategy for Estimating Model Discrimination Power}

\indent\indent
To determine the discrimination power achieved by all these
ratios, we adopt an experimental point of view.
We will choose a particular input boundary condition scenario and particular
values of $\mhalf$ and $\tanb$ as `nature's choice'.
The resulting model will predict certain $\mha$ and $\mcpmone$ values,
which will be measured with small errors.  The same values
for these two observable masses can only be obtained for very specific
$\mhalf$ and $\tanb$ values in any other boundary condition scenario.
Once, the $(\mhalf,\tanb)$ location in each scenario
that yields the observed $\mha$ and $\mcpmone$ is established,
we compute the predictions for all the ratios 
of branching fractions. We use the notation $R_i$, with $i$
specifying any particular ratio; the values of the $R_i$
for the input scenario will be denoted by $R_i^0$.
We also compute the $1\sigma$ error in the measurement of each of these ratios
(denoted $\Delta R_i$) as found assuming that the input model
is nature's choice. We may then compute the expected
$\Delta\chi^2$ for any of the other models relative to the input model as:
\begin{equation}
\Delta\chi^2=\sum_i \Delta\chi_i^2\,,~~{\rm with}~~\Delta\chi_i^2
={(R_i-R_i^0)^2\over \Delta R_i^2}\,.
\label{chisqdef}
\end{equation}
We will see that very large $\Delta\chi^2$ values are typically associated
with an incorrect choice of model.

It is important to note that many other observables
that discriminate between models
will be available from other experimental observations.
An additional $\Delta\chi^2$ contribution
should be added for each observable in assessing the overall
improbability of a model other than the correct one.
However, there are advantages to restricting oneself to the
branching fraction ratios only.
For example, $\mslepr$ 
(which will be readily measured in slepton pair production)
differs substantially at fixed $\mha,\mhalf$ as one moves
between the NS, D and HS scenarios, and would readily distinguish
between the models.  However, $\mslepr$ is primarily
sensitive to the value of the slepton $m_0$ at $\mgut$,
which could differ from the $m_0$ associated with the Higgs fields
if the GUT boundary conditions are nonuniversal.  In contrast,
the branching fraction ratios are primarily sensitive to the Higgs
$m_0$ value relative to $\mhalf$.  Different sets of observables
will have maximial sensitivity to different subsets of the GUT scale
boundary conditions. The Higgs branching fraction ratios should be
very valuable in sorting out the correct relation between $\mhalf$
and the $m_0$ for the Higgs fields, and in determining $\tanb$.

\subsection{A Test Case}

\indent\indent
As a specific example, 
suppose the correct model is \DM\ with $\mhalf=201.7\gev$ and 
$\tanb=7.50$. This would imply $\mha=349.7\gev$, $\mcpmone =149.5\gev$.
The $\mhalf$ and $\tanb$ values required in order
to reproduce these same $\mha$ and $\mcpmone$ values
in the other scenarios are listed in Table~\ref{mhalftanbtable}.
Also given in this table are the predicted values of $\mhh$
and $\mslepr$ for each scenario.
In order to get a first feeling for event numbers and 
for the errors that might be expected for the ratios of interest, we give in
Table~\ref{ratestable} the numbers of events, $\caln$ and $\cald$, predicted
in each scenario for use in determining
the numerators and denominators of Eqs.~(\ref{hhhasusy})-(\ref{hazhl})
and Eqs.~(\ref{hphmsusy})-(\ref{hpwhl}), assuming
$\leff=80\fbi$ at $\rts=1\tev$. These
numbers include the SUSY branching fractions, $\breff$
of Eq.~(\ref{breffdef}), and so forth following
the itemized list of factors given 
earlier.\footnote{Because $\br(\ha\to t\anti t)=0$
and $\br(\hh\to t\anti t)$ is typically small 
for the test case choice of $\mha=349.7\gev$ (given $\mt=175\gev$),
the ratio of Eq.~(\ref{hhhattbb}) and its
numerator event rate are both small. Note that
$\breff(\ha\to b\anti b+t\anti t)=\br(\ha\to b\anti b)$
and that $\breff(\hh\to b\anti b+t\anti t)$ is not
very different from $\br(\hh\to b\anti b)$ for this same reason.}

\begin{table}[hbt]
\caption[fake]{We tabulate the values of $\mhalf$ (in GeV)
and $\tanb$ required in each of our six scenarios in order
that $\mha=349.7\gev$ and $\mcpmone=149.5\gev$.
Also given are the corresponding values of $\mhh$
and $\mslepr$. Masses are in GeV.}
\begin{center}
\begin{tabular}{|c|c|c|c|c|c|c|}
\hline
 & \DM\ & \DP\ & \NSM\ & \NSP\ & \HSM\ & \HSP\ \\
\hline
\hline
$\mhalf$ & 201.7 & 174.4 & 210.6 & 168.2 & 203.9 & 180.0 \\
$\tanb$ & 7.50 & 2.94 & 3.24 & 2.04 & 12.06 & 3.83 \\
$\mhh$ & 350.3 & 355.8 & 353.9 & 359.0 & 350.1 & 353.2 \\
$\mslepr$ & 146.7 & 127.5 & 91.0 & 73.9 & 222.9 & 197.4 \\
\hline
\end{tabular}
\end{center}
\label{mhalftanbtable}
\end{table}

\begin{table}[hbt]
\caption[fake]{We give the numbers of events predicted
in each scenario at the parameter
space locations specified in Table~\ref{mhalftanbtable}
available for determining the numerators and denominators of 
Eqs.~(\ref{hhhasusy})-(\ref{hazhl}) and Eqs.~(\ref{hphmsusy})-(\ref{hpwhl}).
These event rates are those for $\leff=80\fbi$ at $\rts=1\tev$. 
They include all branching fractions.
Our notation is $\caln_{(\#)}$ and $\cald_{(\#)}$ 
for the event rates in the numerator and denominator, respectively,
of the ratio defined in Eq.~(\#).
}
\begin{center}
\begin{tabular}{|c|c|c|c|c|c|c|}
\hline
 & \DM\ & \DP\ & \NSM\ & \NSP\ & \HSM\ & \HSP\ \\
\hline
\hline
$\caln_{(\ref{hhhasusy})}$ & 97.0 & 92.3 & 88.3 & 49.2 & 76.1 & 124.0 \\
$\caln_{(\ref{hhhattbb})}$ & 0.1 & 0.7 & 3.8 & 1.02 & 0.0 & 0.2 \\
$\caln_{(\ref{hhhlhl})}$   & 16.4 & 2.7 & 46.6 & 1.47 & 3.8 & 2.4 \\
$\caln_{(\ref{hazhl})}$ & 2.0 & 1.3 & 9.2 & 0.6 & 0.4 & 1.1 \\
$\cald_{(\ref{hhhasusy})}$ 
      & 198 & 9.6 & 62.1 & 2.6 & 250 & 18.2 \\
$\cald_{(\ref{hhhattbb})-(\ref{hazhl})}$ 
      & 198 & 8.9 & 58.3 & 1.6 & 250 & 18.0 \\
$\caln_{(\ref{hphmsusy})}$ & 225 & 189 & 138 & 135 & 189 & 262 \\
$\caln_{(\ref{hptaunu})}$ & 58.4 & 4.2 & 6.5 & 1.1 & 90.0 & 9.5 \\
$\caln_{(\ref{hpwhl})}$ & 13.0 & 12.8 & 21.9 & 9.0 & 3.3 & 12.3 \\
$\cald_{(\ref{hphmsusy})-(\ref{hpwhl})}$ 
      & 317 & 415 & 445 & 465 & 320 & 348 \\
\hline
\end{tabular}
\end{center}
\label{ratestable}
\end{table}

From Table~\ref{ratestable},
we observe that the $\cald_{(\ref{hhhasusy})-(\ref{hazhl})}$
event rates for the $\mu>0$ scenarios are all rather small
as compared to the event rates for the $\mu<0$ scenarios.
(This happens because the $\mhalf$ and $\tanb$ values required for 
$\mha=349.7\gev$ and $\mcpmone=149.5\gev$ when $\mu>0$
are very close to the scenario boundary.)
For example, if the \DM\ model is nature's choice,
the $\hh\ha$-pair denominator rates
would be $\sim 198$, implying a statistical error of only $\sim\pm 14$.
Assuming systematic error of order 10\%, the net error 
in event number would certainly be $\lsim 35$, \ie\ many $\sigma$
away from any of the $\mu>0$ scenario predictions.  
We also see significantly larger numerator rates $\caln_{(\ref{hhhlhl})}$ 
and $\caln_{(\ref{hptaunu})}$ for the $\mu<0$ scenarios than for
the $\mu>0$ scenarios.
Thus, in this particular case, even before examining the branching fraction
ratios, the $\mu>0$ scenarios could be excluded. 

The $\caln$ and $\cald$ event numbers of Table~\ref{ratestable}
also make apparent the accuracy with which the ratios of
Eqs.~(\ref{hhhasusy})-(\ref{hazhl})
and Eqs.~(\ref{hphmsusy})-(\ref{hpwhl}) can be measured. For example,
the event numbers $\caln_{\ref{hhhasusy}}$ and $\cald_{\ref{hhhasusy}}$ 
show that good statistical precision, $\sim \pm 10\%-15\%$,
can be expected for the ratio of Eq.~(\ref{hhhasusy})
in the $\mu<0$ scenarios. Such statistical precision implies
that this ratio will also clearly distinguish between the input \DM\
scenario and any of the $\mu>0$ model predictions.

\begin{figure}[[htbp]
\vskip 1in
\let\normalsize=\captsize   
\begin{center}
\centerline{\psfig{file=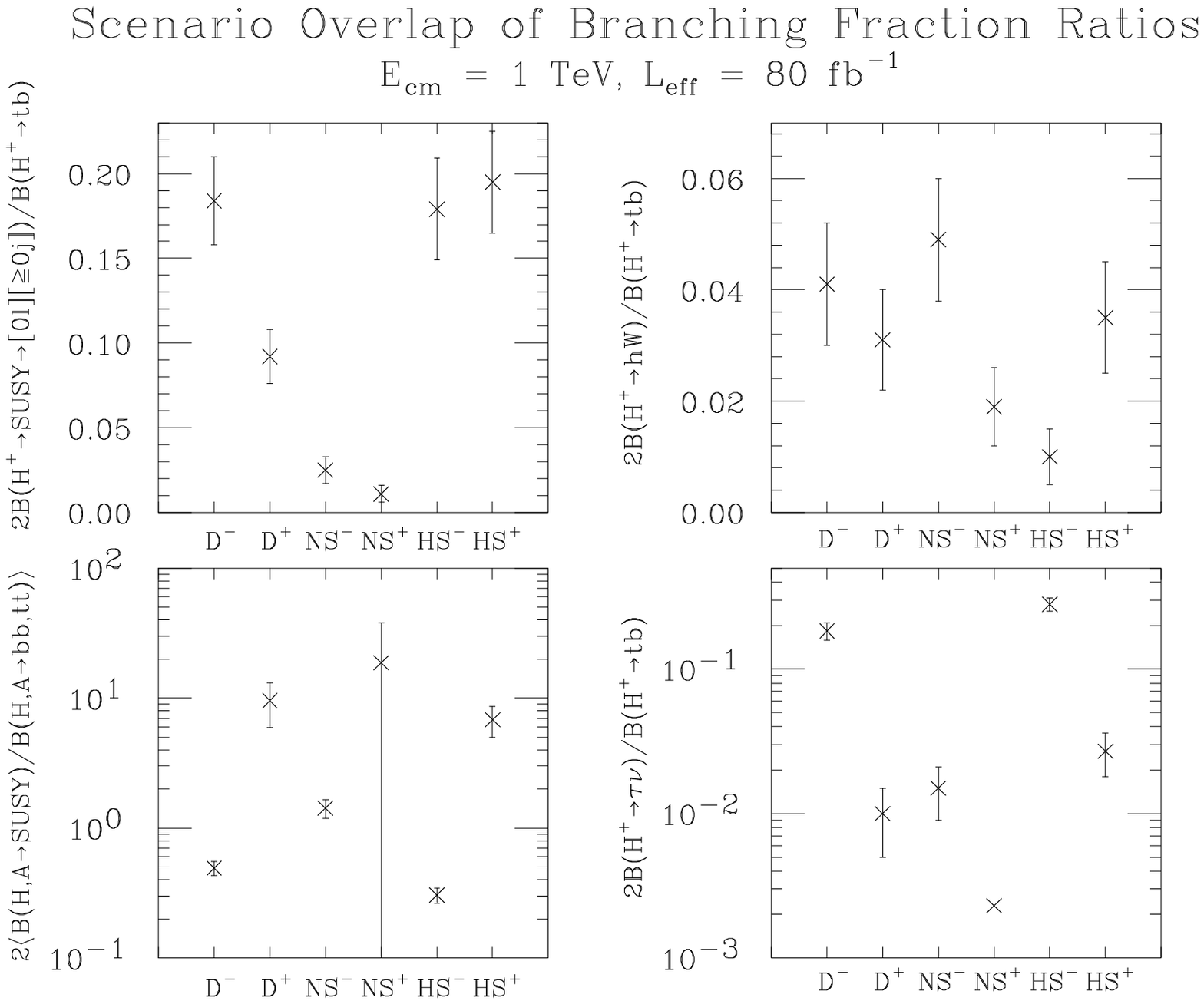,width=12.5cm}}
\smallskip
\begin{minipage}{12.5cm}       
\caption{
We plot the branching fraction ratios $\br(\hp\to\tau^+\nu)/
\br(\hp\to t\anti b)$,
$\br(\hp\to SUSY\to [0\ell][\geq0j])/\br(\hp\to t\anti b)$, 
$\br(\hh ,\ha\to SUSY)/\br_{\rm eff}(\hh ,\ha\to b\anti b,t\anti t)$ and 
$\br(\hp\to\hl W^+)/\br(\hp\to t\anti b)$ with
$\pm 1\sigma$ error bars as a function of scenario, adjusting
$\mhalf$ and $\tanb$ in each scenario so that 
$\mha=349.7\gev$ and $\mcpmone=149.5\gev$ are held fixed. Error bars are
for $\leff=80\fbi$ at $\protect\rts=1\tev$, and are those that
would arise if the input (nature's choice) scenario is that
listed on the horizontal axis. No error bar is shown 
for the $\tau^+\nu/t\anti b$ ratio in the \NSP\ scenario
since the predicted rate is less than 4 events; a very large
error bar should be assumed.
}
\label{fexclusion}
\end{minipage}
\end{center}
\end{figure}

To illustrate the value of the branching fraction ratios more clearly, 
we present in Fig.~\ref{fexclusion} 
a plot which gives the expected values
and the $\pm1\sigma$ errors as a function of scenario
for four of the ratios that will be useful in distinguishing
between the different scenarios at the given input (measured)
values of $\mha$ and $\mcpmone$. In this plot, the errors 
as a function of scenario are those
that are expected if the scenario listed on the horizontal axis
is the correct one. Thus, if the correct model is \DM, the 
central value and $\pm1\sigma$
upper and lower limits for each ratio are those given above the \DM\ 
scenario label on the $x$-axis.
The ability of each ratio to discriminate
between a given scenario on the horizontal axis
and one of the five alternatives is indicated by the extent
to which the $\pm1\sigma$ error bars for the given scenario do
not overlap the central points for the other scenario.
Referring to Fig.~\ref{fexclusion} we observe the following.
\begin{itemize}
\item
The ratio $\br(\hp\to SUSY\to [0\ell][\geq0j])/\br(\hp\to t\anti b)$
succeeds in distinguishing the
\DM\ scenario from all but the \HSM\ and \HSP\ scenarios. 
\item
The ratio $\br(\hp\to\tau^+\nu)/\br(\hp\to t\anti bb)$ provides excellent
discrimination between the \DM\ input scenario and 
the \DP, \NSM, \NSP, and \HSP\
scenarios, all of which must have $\tanb< 4$ (in order to
reproduce $\mha=349.7\gev,\mcpmone=149.5\gev$) as compared to $\tanb=7.5$
for the \DM\ scenario. The much smaller $\tanb$ values imply
much smaller $\tau^+\nu/t\anti b$ ratios, as
was illustrated in Fig.~\ref{taunutbratio}. The more limited
ability of this ratio to discriminate between 
the high $\tanb$ values of 7.5 for \DM\ vs. 12
for \HSM\ is also apparent from Fig.~\ref{taunutbratio}.
\item
The ratio $\br(\hh ,\ha\to SUSY)/\br(\hh ,\ha\to b\anti b,t\anti t)$ 
will strongly rule out $\mu>0$ scenarios if $\mu<0$ is nature's choice.
Due to the small error bars, this ratio provides some discrimination between
the \DM\ and \HSM\ scenarios even though
the predicted central values are not very different.
\item
The ratio $\br(\hp\to\hl \wp)/\br(\hp\to t\anti b)$ is
quite different for the \DM, \NSM, and \HSP\ scenarios
as compared to the \DP, \NSP, and \HSM\ scenarios.  However,
discrimination power is limited by the relatively large error bars.
Nonetheless, this ratio yields a bit more than $2.7\sigma$
discrimination against the \HSM\ model if the \DM\ model is nature's choice.
\end{itemize}
The quite substantial dependence of the ratios on scenario
and location in parameter space, as displayed in 
Figs.~\ref{fhhhasusy}, \ref{fhhhattbb}, \ref{fhhhlhl}, \ref{fhazhl},
\ref{fhphmsusy}, \ref{fhptaunu}, and \ref{fhpwhl},
suggests that similar discrimination will be possible for most input
scenario and parameter space location choices.

\begin{table}[hbt]
\caption[fake]{We tabulate $\Delta\chi^2_i$, see Eq.~(\ref{chisqdef}),
(relative to the \DM\ scenario) for the indicated branching
fraction ratios as a function of scenario,
assuming the measured $\mha$ and $\mcpmone$ values are $349.7\gev$
and $149.5\gev$, respectively. The SUSY channels have been resolved into 
final states involving a fixed number of leptons.  
The error used in calculating each $\Delta\chi^2_i$ is the approximate
$1\sigma$ error (as defined in text)
with which the given ratio $R_i$ could be measured
for $\leff=80\fbi$ at $\rts=1\tev$ {\it assuming that the \DM\
scenario is the correct one}.
}
\begin{center}
{\footnotesize
\begin{tabular}{|c|c|c|c|c|c|c|}
\hline
Ratio & \DM\ & \DP\ & \NSM\ & \NSP\ & \HSM\ & \HSP\ \\
\hline
$2\langle\br(\hh ,\ha\rta SUSY\rta [0\ell][\geq0 j])/$ 
& 0 & 12878 & 1277 & 25243 & 0.77 & 10331 \\
$\breff(\hh ,\ha\rta b \anti b,t\anti t)\rangle$ & & & & & & \\
$2\langle\br(\hh ,\ha\rta SUSY\rta [1\ell][\geq0 j])/$ 
& 0  & 13081 & 2.41 & 5130 & 3.6 & 4783 \\
$\breff(\hh ,\ha\rta b \anti b,t\anti t)\rangle$ & & & & & & \\
$2\langle\br(\hh ,\ha\rta SUSY\rta [2\ell][\geq0j])/$ 
& 0  & 4543 & 5.12 & 92395 & 26.6 & 116 \\
$\breff(\hh ,\ha\rta b \anti b,t\anti t)\rangle$ & & & & & & \\
$\br(\hh\rta \hl\hl)/\br(\hh\rta b\anti b)$ 
& 0 & 109 & 1130 & 1516 & 10.2 & 6.2 \\
$2\br(\hp\rta SUSY\rta [0\ell][\geq0j])/$ 
& 0 & 12.2 & 36.5 & 43.2 & 0.04 & 0.2 \\
$\br(\hp\rta t \anti b)$ & & & & & & \\
$2\br(\hp\rta SUSY\rta [1\ell][\geq0j])/$ 
& 0 & 1.5 & 0.3 & 0.1 & 5.6 & 0.06  \\
$\br(\hp\rta t \anti b)$ & & & & & & \\
$2\br(\hp\rta\hl W)/\br(\hp\rta t\anti b)$ 
& 0 & 0.8 & 0.5 & 3.6 & 7.3 & 0.3 \\
$2\br(\hp\rta\tau\nu )/\br(\hp\rta t\anti b)$ 
& 0 & 43.7 & 41.5 & 47.7 & 13.7 & 35.5 \\
\hline
$\sum_i\Delta\chi^2_i$ & 0 & 30669 & 2493 & 124379 & 68 & 15272 \\
\hline
\end{tabular}
}
\end{center}
\label{chisqtable}
\end{table}

In Table~\ref{chisqtable} we more thoroughly quantify the process of excluding
the \DP, \NSM, \NSP, \HSM, and \HSP\ scenarios relative to the input
\DM\ scenario.  There we give the contribution to $\Delta\chi^2$
(computed relative to the assumed-to-be-correct 
\DM\ scenario) for each of a selection
of independently measurable ratios. Also given for each
of the incorrect scenarios is the sum of these contributions.
This table shows that the \DM\ scenario can be distinguished
from the \DP, \NSM, \NSP,  and \HSP\ scenarios
at an extremely high statistical level.
Further, even though no one of the branching fraction ratios
provides an absolutely clear discrimination between the
\DM\ and the \HSM\ scenarios, the accumulated discrimination power
obtained by considering all the ratios is very substantial. In particular,
although the ratios of Eq.~(\ref{hhhlhl}), (\ref{hazhl}),
and (\ref{hpwhl}) are only poorly measured for $\leff=80\fbi$, 
their accumulated $\Delta\chi^2$ weight can be an important
component in determining the likelihood of a given model and thereby
ruling out incorrect model choices.

Thus, consistency of all the ratios with one another
and with the measured $\mha$, neutralino and chargino
masses will generally restrict the allowed models
to ones that are very closely related.  
The likelihood or probability associated with the best fit to all these 
observables in a model that differs significantly from
the correct model would be very small.  

\subsection{Separating Different SUSY Decay Modes}

\indent\indent
An important issue is the extent to which one can be
sensitive to the branching fractions
for different types of SUSY decays of the Higgs bosons,
relative to one another and relative to the overall SUSY decay
branching fraction. Interesting SUSY decay rates include:
\begin{itemize}
\item
$\br(\hh,\ha\to \cnone\cnone+\snu\snu)$, leading to a totally invisible
final state;
\item
$\br(\hh,\ha\to \slep^+\slep^-)$, where $\slep^{\pm}\to \ell^{\pm} \cnone$ 
or $\nu \cpmone$;
\item
$\br(\hh,\ha\to \cpone\cmone)$, 
where $\cpmone\to \ell^{\pm}\nu\cnone$, $jj\cnone$ or $\slep^{\pm}\snu$ (with
$\slep^{\pm}\snu\to \ell^{\pm}\cnone\nu\cnone$);
\item
$\br(\hpm\to \slep^{\pm}\snu)$, 
where $\slep^{\pm}\to \ell^{\pm}\cnone$, or $\nu\cpmone$;
\item
$\br(\hpm\to \cpmone\cnone)$, 
where $\cpmone\to \ell^{\pm}\nu\cnone$, $jj\cnone$ or $\slep^{\pm}\snu$ (with
$\slep^{\pm}\snu\to \ell^{\pm}\cnone\nu\cnone$).
\end{itemize}
Predictions for such rates depend in a rather detailed fashion
upon the SUSY parameters and would provide valuable information
regarding the SUSY scenario. For example, 
in going from NS to D to HS the masses of the sneutrinos
and sleptons increase relative to those for the charginos and neutralinos.
The $\hh,\ha\to \slep^+\slep^-$ and $\hpm\to \slep^{\pm}\snu$
branching fractions should decline in comparison to 
$\hh,\ha\to\cpone\cmone$ and $\hpm\to \cpmone\cnone$, respectively.
In small sections of the D and NS scenario parameter
spaces, the sleptons and sneutrinos
are sufficiently light that $\cpmone$ decays
almost exclusively to $\slep^{\pm}\snu$ followed by 
$\slep^{\pm}\snu\to \ell^{\pm}\cnone\nu\cnone$, implying that
$\cpmone$ decays would mainly yield leptons and not jets.

The difficulty is that several different SUSY channels can contribute
to any given final state. Two examples were noted earlier:
the $\ell^+\ell^-+\etmiss$
channel receives contributions from both 
$\hh,\ha\to \slep^+\slep^-$ and $\cpone\cmone$
decays; and the $\ell^{\pm}+\etmiss$ channel receives contributions
from $\hpm\to \slep^{\pm}\snu$ and $\cpmone\cnone$.  Another example,
is the purely invisible
$\hh$ or $\ha$ final state; it can arise from either 
$\cnone\cnone$ or $\snu\snu$ (with $\snu\to \nu \cnone$) production. Thus,
the physically distinct channels,
defined by the number of leptons and jets 
present,\footnote{The totally invisible final state would be 
$[0\ell][0j]$, and so forth.} typically have
multiple sources.   Still, a comparison between the rates
for the final states so-defined might be quite revealing.
For instance, if $\cpmone\to \slep^{\pm}\snu$
is not kinematically allowed, the $\cpone\cmone$ final states are expected 
to yield more $1\ell +2j$ and $0\ell+4j$
events than $2\ell+0j$ events, whereas $\slep^+\slep^-$
events will yield only $2\ell+0j$ events. Further, the $\ell$'s
must be of the same type in this latter case.
The effective branching fraction
for $\cpone\cmone\to\ell^+\ell^-+\etmiss$ with both $\ell$'s
of the same type is only 1/81.
In addition, the $\ell$'s in the latter derive from three-body decays
of the $\cpmone$, and would be much softer on average than $\ell$'s
from $\slep^+\slep^-$. Even if this difference is difficult
to see directly via distributions, it will lead to higher
efficiency for picking up the $\slep^+\slep^-$ events.
Of course, if event numbers are 
sufficiently large (which in general they are not) that detailed
kinematical distributions within each final state could be
obtained, they would provide additional information. We do not
pursue this latter possibility here.

Based on the above discussion, the following ratios would appear
to be potentially useful. For the $\hh$ and $\ha$ we consider:
\begin{eqnarray}
&{\br(\hh\to b\anti b)\br(\ha\to [0\ell][0j])
 +\br(\ha\to b\anti b)\br(\hh\to [0\ell][0j])
\over 
  \br(\hh\to b\anti b)\br(\ha\to {\rm SUSY})
 +\br(\ha\to b\anti b)\br(\hh\to {\rm SUSY})
}\;;  & \label{0l0j} \\
&{\br(\hh\to b\anti b)\br(\ha\to [2\ell][0j])
 +\br(\ha\to b\anti b)\br(\hh\to [2\ell][0j])
\over 
  \br(\hh\to b\anti b)\br(\ha\to {\rm SUSY})
 +\br(\ha\to b\anti b)\br(\hh\to {\rm SUSY})
}\;;  & \label{2l0j} \\
&{\br(\hh\to b\anti b)\br(\ha\to [\geq 0\ell][0j])
 +\br(\ha\to b\anti b)\br(\hh\to [\geq 0\ell][0j])
\over 
  \br(\hh\to b\anti b)\br(\ha\to {\rm SUSY})
 +\br(\ha\to b\anti b)\br(\hh\to {\rm SUSY})
}\;;  & \label{geq0l0j} \\
&{\br(\hh\to b\anti b)\br(\ha\to [0\ell][\geq 1j])
 +\br(\ha\to b\anti b)\br(\hh\to [0\ell][\geq 1j])
\over 
  \br(\hh\to b\anti b)\br(\ha\to {\rm SUSY})
 +\br(\ha\to b\anti b)\br(\hh\to {\rm SUSY})
}\;;  & \label{0lgeq1j} \\
&{\br(\hh\to b\anti b)\br(\ha\to [1\ell][\geq 1j])
 +\br(\ha\to b\anti b)\br(\hh\to [1\ell][\geq 1j])
\over 
  \br(\hh\to b\anti b)\br(\ha\to {\rm SUSY})
 +\br(\ha\to b\anti b)\br(\hh\to {\rm SUSY})
}\;;  & \label{1lgeq1j}
\end{eqnarray}
(As before, $\mha\sim\mhh$ implies that we cannot separate the
$\hh$ and $\ha$ via the tagging 
procedure.\footnote{The $\ha\to\slep^+\slep^-$ branching
ratio turns out to be rather small in the three GUT scenarios
studied --- the required $L-R$ mixing is numerically very
small in the slepton sector.}) 
Once again, we employ shorthand notations for the quantities appearing in
Eqs.~(\ref{0l0j})-(\ref{1lgeq1j}). For example,
the ratio of Eq.~(\ref{0l0j}) will be denoted by
\begin{equation}
{\langle\br(\ha,\hh\to [0\ell][0j])\br(\hh,\ha\to b\anti b)\rangle\over 
\langle\br(\ha,\hh\to {\rm SUSY})\br(\hh,\ha\to b\anti b)\rangle}\,
~~~{\rm or}~~~\left.{\br(\hh,\ha\to [0\ell][0j])\over
\br(\hh,\ha\to {\rm SUSY})}\right|_{\rm eff}
\end{equation}
in what follows.

For the $\hpm$ we consider the ratios:
\begin{eqnarray}
& {\br(\hp \to [1\ell][0j])\br(\hm\to b\anti t)
  +\br(\hm\to[1\ell][0j])\br(\hp\to t\anti b)
\over 
\br(\hp\to {\rm SUSY})\br(\hm\to b\anti t)+
\br(\hm\to {\rm SUSY})\br(\hp\to t\anti b)} 
\;;  & \label{1l0jhp} \\
& {\br(\hp\to [\geq1\ell][0j])\br(\hm \to b\anti t)
  +\br(\hm\to [\geq1\ell][0j])\br(\hp \to t\anti b)
\over 
\br(\hp\to {\rm SUSY})\br(\hm\to b\anti t)+
\br(\hm\to {\rm SUSY})\br(\hp\to t\anti b)}
\;;  & \label{geq1l0jhp} \\
& {\br(\hp\to [0\ell][\geq1 j])\br(\hm \to b\anti t)
  +\br(\hm\to [0\ell][\geq1 j])\br(\hp \to t\anti b)
\over 
\br(\hp\to {\rm SUSY})\br(\hm\to b\anti t)+
\br(\hm\to {\rm SUSY})\br(\hp\to t\anti b)}
\;.  & \label{0lgeq1jhp} 
\end{eqnarray}
The ratios of Eqs.~(\ref{1l0jhp})-(\ref{0lgeq1jhp}) reduce to:
\begin{equation}
{\br(\hp\to [1\ell][0j])\over \br(\hp\to {\rm SUSY})}\,,~~
{\br(\hp\to [\geq 1\ell][0j])\over \br(\hp\to {\rm SUSY})}\,,~~
{\br(\hp\to [0\ell][\geq 1j])\over \br(\hp\to {\rm SUSY})}\,,
\end{equation}
respectively.

Also of interest are ratios of the different numerator terms to one
another within the above neutral and charged Higgs boson sets.
All the ratios that one can form
have the potential to provide important tests of
the Higgs decays to the supersymmetric particle pair final states. 

\begin{figure}[[htbp]
\vskip 1in
\let\normalsize=\captsize   
\begin{center}
\centerline{\psfig{file=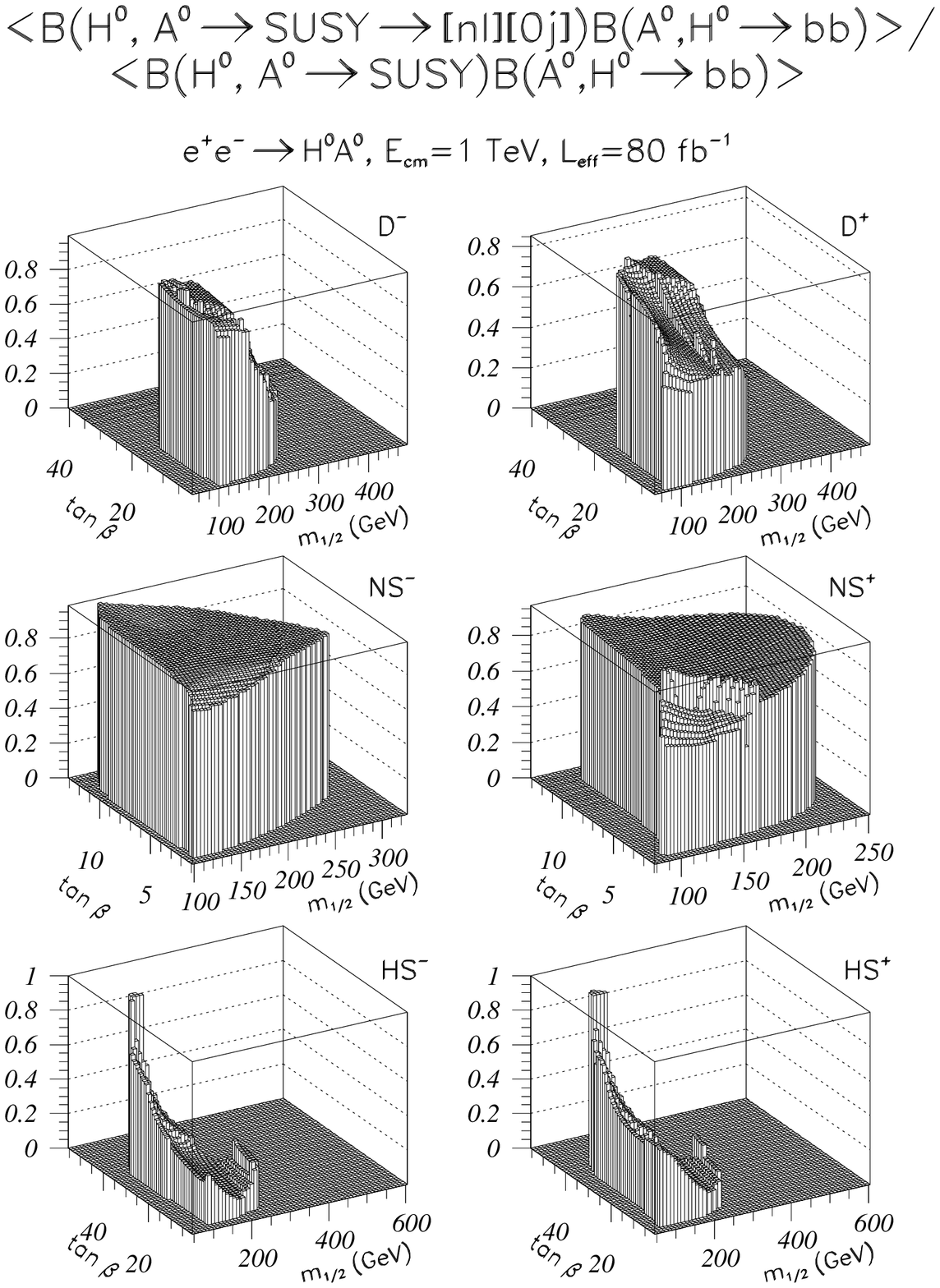,width=10.5cm}}
\smallskip
\begin{minipage}{12.5cm}       
\caption{
We present lego plots of the ratio of Eq.~(\ref{geq0l0j}) 
in each of the six scenarios as a function of location in
$(\mhalf,\tanb)$ parameter space. 
The value of the ratio is given by the height on the $z$-axis.
Non-zero values of the ratio are given only in regions where
there are at least 4 events in the numerator
after including tagging/reconstruction efficiencies.
}
\label{fgeq0l0j}
\end{minipage}
\end{center}
\end{figure}

\begin{figure}[[htbp]
\vskip 1in
\let\normalsize=\captsize   
\begin{center}
\centerline{\psfig{file=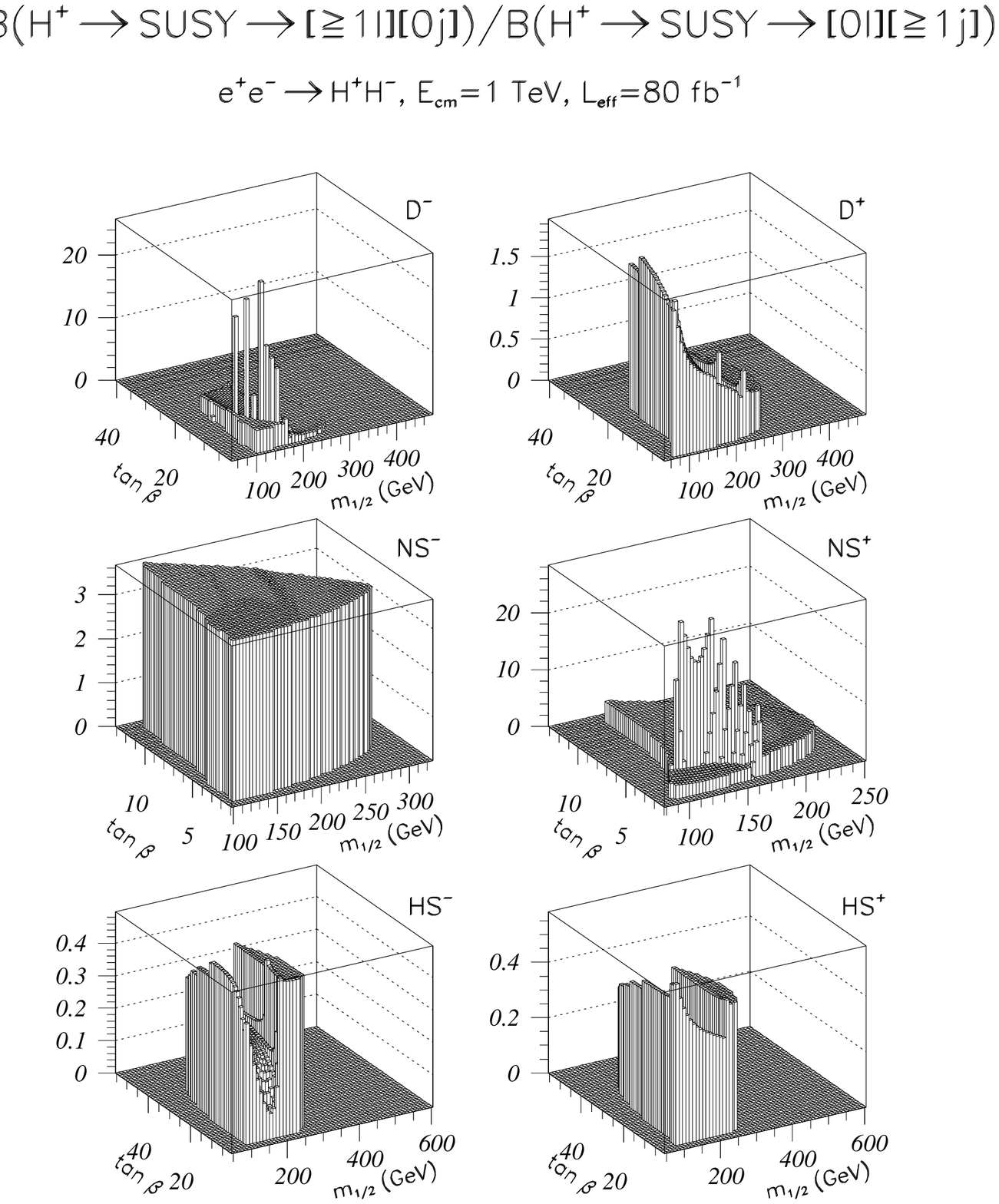,width=10.5cm}}
\smallskip
\begin{minipage}{12.5cm}       
\caption{
We present lego plots of the numerator of Eq.~(\ref{geq1l0jhp}) 
divided by the numerator of Eq.~(\ref{0lgeq1jhp})
in each of the six scenarios as a function of location in
$(\mhalf,\tanb)$ parameter space. 
The value of the ratio is given by the height on the $z$-axis.
Non-zero values of the ratio are given only in regions where
there are at least 4 events in both the numerator and denominator
after including tagging/reconstruction efficiencies.
}
\label{fgeq1l0jhp}
\end{minipage}
\end{center}
\end{figure}

To illustrate, we present two figures. In Fig.~\ref{fgeq0l0j} 
we present three-dimensional lego plots of the ratio
of Eq.~(\ref{geq0l0j}) as a function of location in
$(\mhalf,\tanb)$ parameter space. (Because of the combination
of slow variation and very sharp changes, the contour plots
similar to those presented earlier are rather difficult to interpret.)
In Fig.~\ref{fgeq1l0jhp}, we plot the numerator
of Eq.~(\ref{geq1l0jhp}) divided by the numerator of Eq.~(\ref{0lgeq1jhp}).
In both sets of lego plots, the ratio is set to zero if there are fewer than
4 events in the numerator or denominator
after including the earlier-discussed tagging/reconstruction efficiencies
and assuming $\protect\rts=1\tev$ and $L_{\rm eff}=80\fbi$.

The most important feature apparent from these figures is the generally
decreasing magnitude of these two ratios as one moves from the NS to the D to 
the HS scenario.  This is a result of the decreasing importance of
slepton/sneutrino-related decays as compared to chargino/neutralino-based
decays. When the latter types of decay are prevalent, a much
larger fraction of the events will have jets than if the former decays
dominate.  The decreasing importance of the slepton/sneutrino class
is to be expected due to the increasing mass of these states as
$m_0$ increases in going from NS to D to HS.
The occasionally very large values of 
$\br(\hp\to [\geq 1\ell][0j])/\br(\hp\to [0\ell][\geq 1j])$
in Fig.~\ref{fgeq1l0jhp} in the \DM\ and \NSP\
plots occur in the small wedges of parameter
space where $\cpmone\to \slep^{\pm}\snu$
decays are kinematically allowed, and final states containing
only jets must arise from higher ino states and, thus, are very rare.

It should be apparent from these two figures that rather dramatic
differences between the scenarios at a given $(\mhalf,\tanb)$ location
are the norm.  
In general, statistics are such that the different
scenarios can be distinguished from one another at a substantial
level of significance just on the basis of these two ratios.
Ratios other than the two plotted ones can also
provide good discrimination. We shall illustrate this for
our standard $\mha=349.7\gev,\mcpmone=149.5\gev$
point discussed in association with Tables~\ref{mhalftanbtable}
and \ref{ratestable}, Fig.~\ref{fexclusion} and Table~\ref{chisqtable}.

Table~\ref{mhalftanbtable} gives the $(\mhalf,\tanb)$ parameters
required for $\mha=349.7\gev,\mcpmone=149.5\gev$ in each of
the six GUT scenarios.  In Table~\ref{susyratestable} 
the event rates for the SUSY final states corresponding 
to the numerators of the ratios listed
in Eqs.~(\ref{0l0j})-(\ref{1lgeq1j}) and (\ref{1l0jhp})-(\ref{0lgeq1jhp})
are given for these $(\mhalf,\tanb)$ values. 
We will follow the
same notation in terms of $\caln_{\rm (Eq.~\#)}$ as 
for Table~\ref{ratestable}.
An examination of Table~\ref{susyratestable} reveals event rates in 
the individual channels that vary from a few events, implying
poor statistics, to 50 or 60 events, for which statistical accuracy
would be quite reasonable.

\begin{table}[hbt]
\caption[fake]{For the $(\mhalf,\tanb)$ values required
for $\mha=349.7\gev,\mcpmone=149.5\gev$, 
we tabulate the numbers of events predicted
in each scenario in the final states corresponding to
the numerators and denominators of Eqs.~(\ref{0l0j})-(\ref{1lgeq1j}) and 
(\ref{1l0jhp})-(\ref{0lgeq1jhp}).
These rates are those obtained for $\leff=80\fbi$ at $\rts=1\tev$. 
They include all branching fractions.
}
\begin{center}
\begin{tabular}{|c|c|c|c|c|c|c|}
\hline
 & \DM\ & \DP\ & \NSM\ & \NSP\ & \HSM\ & \HSP\ \\
\hline
\hline
$\caln_{(\ref{0l0j})}$ & 14.8 & 20.4 & 64.3 & 8.7 & 7.7 & 14.7 \\
$\caln_{(\ref{2l0j})}$ & 29.5 & 20.4 & 15.6 & 19.5 & 1.4 & 6.8 \\
$\caln_{(\ref{geq0l0j})}$ & 53.7 & 43.3 & 79.8 & 30.2 & 9.1 & 21.7 \\
$\caln_{(\ref{0lgeq1j})}$ & 10.8 & 9.8 & 3.1 & 3.0 & 30.5 & 37.2 \\
$\caln_{(\ref{1lgeq1j})}$ & 10.8 & 19.3 & 1.8 & 3.4 & 5.6 & 22.1 \\
$\cald_{(\ref{0l0j})-(\ref{1lgeq1j})}$ 
& 97.2 & 87.9 & 86.4 & 37.7 & 76.1 & 124 \\
$\caln_{(\ref{1l0jhp})}$ & 26.0 & 24.3 & 40.6 & 40.5 & 13.4 & 25.9 \\
$\caln_{(\ref{geq1l0jhp})}$ & 26.0 & 26.2 & 40.6 & 43.5 & 13.4 & 25.9 \\
$\caln_{(\ref{0lgeq1jhp})}$ & 58.4 & 38.3 & 11.1 & 5.2 & 57.2 & 67.9 \\
$\cald_{(\ref{1l0jhp})-(\ref{0lgeq1jhp})}$ 
& 225 & 189 & 138 & 135 & 189 & 262 \\
\hline
\end{tabular}
\end{center}
\label{susyratestable}
\end{table}

Not surprisingly,
the ratios of rates of the various SUSY channels can contribute significantly
to our ability to discriminate between different GUT scenarios.
To illustrate, we follow the same procedure as in Table~\ref{chisqtable}.
Taking $\mha=349.7\gev$ and $\mcpmone=149.5\gev$, we assume that
the correct scenario is \DM\ and compute the $\Delta\chi^2$
by which the prediction for a given ratio in the
other scenarios deviates from the \DM\ prediction.  Statistics
are computed on the basis of the expected \DM\ rates, as given in
Table~\ref{susyratestable}. The resulting $\Delta \chi^2$ values
are given in Table~\ref{susychisqtable}.  Since these ratios are not
all statistically independent of one another, we do not sum their
$\Delta\chi^2_i$'s to obtain an overall discrimination level.  However,
a rough indication of the level at which 
any given scenario can be ruled out relative to the \DM\ is obtained if we
add the largest $\Delta\chi^2_i$ from the neutral Higgs
list and the largest from the charged Higgs list. The weakest discrimination
level following this procedure is $\Delta\chi^2\sim 15$ 
in the case of the \DP\ scenario. Note that this scenario is highly
unlikely on the basis of the earlier $\sum_i\Delta\chi^2_i$ value
listed in Table~\ref{chisqtable}.  In Table~\ref{chisqtable}, the weakest
discrimination was that for the \HSM\ scenario with 
$\sum_i\Delta\chi^2_i\sim 68$.  We observe from Table~\ref{susychisqtable}
that the ratio 
$\br(\hh,\ha\to [0\ell][0j])/\br(\hh,\ha\to [2\ell][0j])|_{\rm eff}$
has $\Delta\chi^2_i\sim 928$ for the \HSM\ case, which would certainly
rule it out.

\begin{table}[hbt]
\caption[fake]{We tabulate $\Delta\chi^2_i$, see Eq.~(\ref{chisqdef}),
(relative to the \DM\ scenario) for the indicated
ratios as a function of scenario,
assuming the measured $\mha$ and $\mcpmone$ values are $349.7\gev$
and $149.5\gev$, respectively. The SUSY channels have been resolved into 
final states involving a restricted number of leptons and jets.
Only those ratios with substantial power for discriminating between
scenarios are tabulated.
The error used in calculating each $\Delta\chi^2_i$ is the approximate
$1\sigma$ error (as defined in text)
with which the given ratio $R_i$ could be measured
for $\leff=80\fbi$ at $\rts=1\tev$ {\it assuming that the \DM\
scenario is the correct one}. 
}
\begin{center}
{\footnotesize
\begin{tabular}{|c|c|c|c|c|c|c|}
\hline
Ratio & \DM\ & \DP\ & \NSM\ & \NSP\ & \HSM\ & \HSP\ \\
\hline
$\br(\hh ,\ha\to[0\ell][0j])/\br(\hh ,\ha\to {\rm SUSY})|_{\rm eff}$ 
& 0 & 3.5 & 193 & 3.4 & 1.4 & 0.6 \\
$\br(\hh ,\ha\to[\geq0\ell][0j])/\br(\hh ,\ha\to {\rm SUSY})|_{\rm eff}$ 
& 0 & 0.4 & 15.3 & 6.8 & 20.9 & 15.8 \\
$\br(\hh ,\ha\to[0\ell][0j])/\br(\hh ,\ha\to[2\ell][0j])|_{\rm eff}$ 
& 0 & 9.6 & 503 & 0.1 & 928 & 105 \\
$\br(\hh ,\ha\to[0\ell][0j])/\br(\hh ,\ha\to[\geq0\ell][0j])|_{\rm eff}$ 
& 0 & 5.8 & 41.9 & 0.03 & 48.4 & 24.5 \\
$\br(\hh ,\ha\to[0\ell][0j])/\br(\hh ,\ha\to[0\ell][\geq1j])|_{\rm eff}$ 
& 0 & 1.4 & 1074 & 6.4 & 3.5 & 2.7 \\
$\br(\hh ,\ha\to[0\ell][0j])/\br(\hh ,\ha\to[1\ell][\geq1j])|_{\rm eff}$ 
& 0 & 0.3 & 3520 & 4.3 & 0 & 1.4 \\
$\br(\hp\to[\geq1\ell][0j])/\br(\hp\to {\rm SUSY})$ 
& 0 & 1.0 & 56.2 & 75.2 & 3.4 & 0.5 \\
$\br(\hp\to[0\ell][\geq1j])/\br(\hp\to {\rm SUSY})$ 
& 0 & 2.1 & 21.7 & 33.4 & 1.3 & 0 \\
$\br(\hp\to[\geq1\ell][0j])/\br(\hp\to[0\ell][\geq1j])$ 
& 0 & 5.2 & 930 & 5738 & 4.0 & 0.4 \\
\hline
\end{tabular}
}
\end{center}
\label{susychisqtable}
\end{table}

The above illustrations demonstrate that
the ratios of rates for individual SUSY channels
correlate strongly with the underlying physics of the different GUT scenarios
(light vs. heavy sleptons in particular) and add a powerful component to our
ability to determine the correct scenario.

\section{Summary and Conclusions}

\indent\indent
In this paper, we have considered detecting and studying
the heavy Higgs bosons of the minimal supersymmetric model
when pair produced in $\epem$ or $\mupmum$ collisions.
We have shown that, in the SUSY GUT models studied,
the target luminosities
of $L=200\fbi$ and $L=1000\fbi$ at $\rts=1\tev$ and $4\tev$,
respectively, will allow detection of $\hh\ha$ and $\hp\hm$
pair production throughout essentially all of the model parameter
space which is allowed by theoretical and kinematic constraints,
despite the presence of SUSY decay
modes of the $\hh,\ha,\hpm$ at a significant level.
The all-jet and high-multiplicity
final states coming from $\hh,\ha\to b\anti b,t\anti t$
and $\hp\to t\anti b,\hm\to b\anti t$ are essentially background free and
provide appropriate and efficient signals
with rates that are adequate even when SUSY decays are present. 
In the all-jet channels, the individual
Higgs boson masses, $\mha$, $\mhh$ and $\mhp$, can be measured
and the approximate degeneracy ($\mha\sim\mhh\sim\mhpm$) predicted by the MSSM
can be checked.

Once the Higgs bosons are detected and their masses determined, the relative
branching fractions for the decay of a single Higgs boson
can be measured by `tagging' (\ie\ identifying) one member of the $\hh\ha$
or $\hp\hm$ pair in an all-jet mode, and then looking
at the ratios of the numbers of events in different event classes
on the opposing side.  In this way, the relative branching
ratios of Eqs.~(\ref{hhhasusy})-(\ref{hazhl}),
Eqs.~(\ref{hphmsusy})-(\ref{hpwhl}), Eqs.~(\ref{0l0j})-(\ref{1lgeq1j}),
and Eqs.~(\ref{1l0jhp})-(\ref{0lgeq1j})
can be measured with reasonable accuracy whenever parameters are such
that the final states in the numerator and denominator both
have significant event rate.\footnote{We focus on event rate ratios
rather than the absolute rates in the many different channels since the
possibly large systematic errors of the absolute rates will tend to
cancel in the ratios. In some cases, 
absolute event rates are so different that they would
also provide substantial discrimination
between different models, despite the possibly large
systematic errors.}
We find that the measured Higgs masses and relative branching fractions,
in combination with direct measurements of the chargino and neutralino
masses, will over-constrain and very strongly limit the possible
SUSY GUT models.  

The specific SUSY GUT models considered are moderately
conservative in that they
are characterized by universal boundary conditions. In all,
we delineated expectations for six different models,
requiring correct electroweak symmetry
breaking via evolution from the GUT scale to $\mz$.
For each model, there are only 
two parameters: $\mhalf$ (the universal gaugino) mass; and
$\tanb$ (the usual Higgs field vacuum expectation value ratio).
Each model is characterized by a definite relation of
the universal soft-SUSY-breaking scalar mass, $m_0$, and the universal
mixing parameter, $A_0$, to $\mhalf$, as well as by a choice for
the sign of $\mu$ (the Higgs superfield mixing coefficient).

The strategy for checking the consistency of a given GUT hypothesis
is straightforward. First, the measured $\ha$, neutralino
and chargino masses are, in almost all cases,
already sufficient to determine the $\mhalf$
and $\tanb$ values required in the given GUT scenario with good precision.  
The value of $\tanb$ so obtained should agree with that determined from 
chargino pair production rates. The Higgs sector branching fractions
can then be predicted and become an
important testing ground for the consistency of the proposed GUT hypothesis
as well as for testing the MSSM two-doublet Higgs sector structure per se.

Within the list of ratios of branching fractions given in 
Eqs.~(\ref{hhhasusy})-(\ref{hazhl}) and (\ref{hphmsusy})-(\ref{hpwhl}),
the average\footnote{Only the average can be determined
given that typically $\mha\sim\mhh$.}
$\hh,\ha\to {\rm SUSY}$, the $\hp\to {\rm SUSY}$ and the $\hp\to\taup\nu$
branching fractions typically fix a relatively precise location
in $(\mhalf,\tanb)$ parameter space. These values can be compared
to those required by the $\mha$ and $\mcpmone$ mass measurements. 
Consistency within experimental errors is typically only possible
for a small set of closely related models. In the sample situation
detailed in Section 4, where we assumed that one of the six
GUT models was correct and computed statistical errors
on that basis, only one of the remaining five models could possibly
be confused with the input model after measuring the above three
branching fractions relative to that for the final state used for tagging.
By subdividing the SUSY signal into final states with a definite number
of leptons and any number of jets, and considering as well
the $\hh\to\hl\hl$, $\ha\to Z\hl$
and $\hp\to \wp\hl$ branching fractions, we found it possible to distinguish
between these two choices at a very substantial
statistical level.  Thus, a unique model
among the six rather similar models is singled out by combining
measurements from the Higgs sector with those from conventional SUSY
pair production. In short, measurements deriving from pair production
of Higgs particles can have a great impact
upon our ability to experimentally determine the correct SUSY GUT model.

The above discussion has left aside the fact that
for universal soft-scalar masses the measured value
of the slepton mass would determine
the relative magnitude of $m_0$ and $\mhalf$.  Of the two models
mentioned just above, one has a large $m_0/\mhalf$ value
and the other a much smaller value.  They could be easily distinguished
on the basis of $\mslep$ alone. However, if the soft-scalar
slepton mass is not the same as the soft-scalar Higgs field masses
at the GUT scale, the branching fraction ratios would give the best
indication of the relative size of the soft-scalar Higgs mass
as compared to $\mhalf$.

More information regarding the slepton/sneutrino mass scale
and additional ability to discrminate between models
are both realized by subdividing the SUSY decays of the Higgs bosons 
in a way that is sensitive to the relative branching fractions for
slepton/sneutrino vs. chargino/neutralino decays. Slepton/sneutrino
channels essentially only produce leptons in the final state, whereas
the jet component is typically larger than
the leptonic component for chargino/neutralino decays (other than
the totally invisible $\cnone\cnone$ mode). Thus, we 
are able to define individual
SUSY channels, characterized by a certain number of leptons
and/or jets, which display a strong correlation
with the slepton/sneutrino decay component. We find that these
individual channels have sufficiently large event rates
that the ratios of the branching fractions for these channels can
typically be determined with reasonable statistical precision. 
For the earlier-mentioned input model,
we can compute the statistical level at which the other five GUT scenarios
would be ruled out using various of these ratios of branching fractions.
Excellent discrimination between models on this basis is found.

In conclusion, our study shows that not only will detection
of Higgs pair production in $\epem$ or $\mupmum$
collisions (at planned luminosities) be possible for most of the kinematically accessible
portion of parameter space in a typical GUT model, but also the detailed rates
for and ratios of different neutral and charged Higgs decay final states
will very strongly constrain the choice of GUT-scale boundary conditions.
In estimating experimental sensitivity for Higgs pair detection
and for measuring Higgs masses and branching fractions, we included
substantial inefficiencies and all relevant branching fractions.
Although we believe that our estimates are relatively conservative,
it will be important to re-visit this analysis using a full Monte Carlo
detector simulation.

\section{Acknowledgements}

\indent\indent
This work was supported in part by Department of Energy under
grant No. DE-FG03-91ER40674
and by the Davis Institute for High Energy Physics. We wish
to thank C.H. Chen for making his evolution program available to us.

\clearpage
 

\begin{thebibliography}{99}

\bibitem{langacker} P. Langacker and M. Luo, \prdj{44} (1991) 817;
U. Amaldi, W. de Boer and H. Furstenau, \plbj{260} (1991) 447;
J. Ellis, S. Kelley and D. Nanopoulos, \plbj{260} (1991) 131.

\bibitem{hhg} See 
J.F. Gunion, H.E. Haber, G.L. Kane and S. Dawson,
{\it The Higgs Hunters Guide}, Addison-Wesley Publishing,
and references therein.

\bibitem{dpfreport} J.F. Gunion, A. Stange, and S. Willenbrock,
{\it Weakly-Coupled Higgs Bosons}, preprint UCD-95-28 (1995),
to be published in {\it Electroweak Symmetry Breaking 
and New Physics at the TeV
Scale}, edited by T.L. Barklow, S. Dawson, H.E. Haber, and J.L.
Siegrist (World Scientific, Singapore, 1996).

\bibitem{gkgamgamsusy} J.F. Gunion, J. Kelly, and J. Ohnemus,
\prdj{51} (1994) 2101.

\bibitem{kalzer} {A. Djouadi, J. Kalinowski, P. Ohmann and P.M. Zerwas,
hep-ph/9605339.}


\bibitem{mupmumconferences}
{\it Proceedings of the First Workshop on the Physics
Potential
and Development of $\mu^+\mu^-$ Colliders}, Napa, California (1992), Nucl.\
Instru.\ and Meth.\ {\bf A350}, 24 (1994);
{\it Proceedings of the Second Workshop on the Physics
Potential and Development of $\mu^+\mu^-$ Colliders}, Sausalito, California
(1994), ed.\ by D.~Cline, American Institute of Physics Conference
Proceedings 352;
{\it Proceedings of the 9th Advanced ICFA Beam Dynamics
Workshop: Beam Dynamics and Technology Issues for $\mu^+\mu^-$ Colliders},
Montauk, Long Island, (1995), to be published.


\bibitem{ghino} J.F. Gunion and H.E. Haber, \prdj{37} (1988) 2515.

\bibitem{noscale} For a review and 
references, see A.B. Lahanas and D.V. Nanopoulos, \prepj{145} (1987) 1.

\bibitem{dilaton} A. Brignole, L.E. Ibanez, and C. Munoz, \npbj{422}
(1994) 125, Erratum, \ibid, {\bf B436} (1994) 747.
See also V.S. Kaplunovsky and J. Louis, \plbj{306} (1993) 269.


\bibitem{ganderson} {G. Anderson, D. Castano,
{\it Phys. Rev.} {\bf D52}, 1693-1700 (1995).}

\bibitem{isasugra} {H. Baer, F. Paige, S. Protpopescu, and X. Tata, in
{\it Proceedings of the Workshop On Physics at Current Accelerators and
Supercolliders}, eds. J. Hewett, A. White, and D. Zeppenfeld, Argonne
National Laboratory (1993).}

\bibitem{chen} We thank C.H.~Chen for making his program available
to us.

\bibitem{bgkp} H. Baer, J.F. Gunion, C. Kao, and H. Pois, \prdj{51} (1995)
2159.

\end{thebibliography}
\end{document}